
\magnification=\magstep1

\newbox\SlashedBox
\def\slashed#1{\setbox\SlashedBox=\hbox{#1}
\hbox to 0pt{\hbox to 1\wd\SlashedBox{\hfil/\hfil}\hss}#1}
\def\hboxtosizeof#1#2{\setbox\SlashedBox=\hbox{#1}
\hbox to 1\wd\SlashedBox{#2}}

\def\mathslashed#1{\setbox\SlashedBox=\hbox{$#1$}
\hbox to 0pt{\hbox to 1\wd\SlashedBox{\hfil/\hfil}\hss}#1}

\def\ifsmall{\iffalse}  
\def\titlepagefont{}  

\def\DefineTeXgraphics{%
\special{ps::[global] /TeXgraphics { } def}}  

\def\today{\ifcase\month\or January\or February\or March\or April\or May
\or June\or July\or August\or September\or October\or November\or
December\fi\space\number\day, \number\year}
\def\eatPrefix19{}
\def\Year{\expandafter\eatPrefix\the\year}
\newcount\hours \newcount\minutes
\def\monthname{\ifcase\month\or
January\or February\or March\or April\or May\or June\or July\or
August\or September\or October\or November\or December\fi}
\def\shortmonthname{\ifcase\month\or
Jan\or Feb\or Mar\or Apr\or May\or Jun\or Jul\or
Aug\or Sep\or Oct\or Nov\or Dec\fi}

\def\TimeStamp{\hours\the\time\divide\hours by60%
\minutes -\the\time\divide\minutes by60\multiply\minutes by60%
\advance\minutes by\the\time%
${\rm \shortmonthname}\cdot\if\day<10{}0\fi\the\day\cdot\the\year%
\qquad\the\hours:\if\minutes<10{}0\fi\the\minutes$}







\newif\ifdraftmode
\newif\ifleftlabels  

\def\nolabels{\def\wrlabeL##1{}\def\eqlabeL##1{}\def\reflabeL##1{}}
\def\writelabels{\def\wrlabeL##1{\leavevmode\vadjust{\rlap{\smash%
{\line{{\escapechar=` \hfill\rlap{\sevenrm\hskip.03in\string##1}}}}}}}%
\def\eqlabeL##1{{\escapechar-1\rlap{\sevenrm\hskip.05in\string##1}}}%
\def\reflabeL##1{\noexpand\rlap{\noexpand\sevenrm[\string##1]}}}
\def\writeleftlabels{\def\wrlabeL##1{\leavevmode\vadjust{\rlap{\smash%
{\line{{\escapechar=` \hfill\rlap{\sevenrm\hskip.03in\string##1}}}}}}}%
\def\eqlabeL##1{{\escapechar-1%
\rlap{\sixrm\hskip.05in\string##1}%
\llap{\sevenrm\string##1\hskip.03in\hbox to \hsize{}}}}%
\def\reflabeL##1{\noexpand\rlap{\noexpand\sevenrm[\string##1]}}}
\nolabels

\newdimen\fullhsize
\newdimen\hstitle
\hstitle=\hsize 
\newdimen\hsbody
\hsbody=\hsize 
\newdimen\hbodyoffset
\hbodyoffset=\hoffset 
\newbox\leftpage
\def\abstract#1{#1}
\def\rotated{\special{ps: landscape}
\magnification=1000  
\baselineskip=14pt
\global\hstitle=9truein\global\hsbody=4.75truein
\global\vsize=7truein\global\voffset=-.31truein
\global\hoffset=-0.54in\global\hbodyoffset=-.54truein
\global\fullhsize=10truein
\def\DefineTeXgraphics{%
\special{ps::[global]
/TeXgraphics {currentpoint translate 0.7 0.7 scale
              -80 0.72 mul -1000 0.72 mul translate} def}}
\let\lr=L
\def\ifsmall{\iftrue}
\def\titlepagefont{\twelvepoint}
\trueseventeenpoint
\def\almostshipout##1{\if L\lr \count1=1
      \global\setbox\leftpage=##1 \global\let\lr=R
   \else \count1=2
      \shipout\vbox{\hbox to\fullhsize{\box\leftpage\hfil##1}}
      \global\let\lr=L\fi}

\output={\ifnum\count0=1 
 \shipout\vbox{\hbox to \fullhsize{\hfill\pagebody\hfill}}\advancepageno
 \else
 \almostshipout{\leftline{\vbox{\pagebody\makefootline}}}\advancepageno
 \fi}

\def\abstract##1{{\leftskip=1.5in\rightskip=1.5in ##1\par}} }

\def\linemessage#1{\immediate\write16{#1}}

\global\newcount\secno \global\secno=0
\global\newcount\appno \global\appno=0
\global\newcount\meqno \global\meqno=1
\global\newcount\subsecno \global\subsecno=0
\global\newcount\figno \global\figno=0

\newif\ifAnyCounterChanged
\let\terminator=\relax
\def\normalize#1{\ifx#1\terminator\let\next=\relax\else%
\if#1i\aftergroup i\else\if#1v\aftergroup v\else\if#1x\aftergroup x%
\else\if#1l\aftergroup l\else\if#1c\aftergroup c\else%
\if#1m\aftergroup m\else%
\if#1I\aftergroup I\else\if#1V\aftergroup V\else\if#1X\aftergroup X%
\else\if#1L\aftergroup L\else\if#1C\aftergroup C\else%
\if#1M\aftergroup M\else\aftergroup#1\fi\fi\fi\fi\fi\fi\fi\fi\fi\fi\fi\fi%
\let\next=\normalize\fi%
\next}
\def\makeNormal#1#2{\def\doNormalDef{\edef#1}\begingroup%
\aftergroup\doNormalDef\aftergroup{\normalize#2\terminator\aftergroup}%
\endgroup}

\def\warnIfChanged#1#2{%
\ifundef#1
\else\begingroup%
\edef\oldDefinitionOfCounter{#1}\edef\newDefinitionOfCounter{#2}%
\ifx\oldDefinitionOfCounter\newDefinitionOfCounter%
\else%
\linemessage{Warning: definition of \noexpand#1 has changed.}%
\global\AnyCounterChangedtrue\fi\endgroup\fi}

\def\Section#1{\global\advance\secno by1\relax\global\meqno=1%
\global\subsecno=0%
\bigbreak\bigskip
\centerline{\twelvepoint \bf %
\the\secno. #1}%
\par\nobreak\medskip\nobreak}
\def\tagsection#1{%
\warnIfChanged#1{\the\secno}%
\xdef#1{\the\secno}%
\ifWritingAuxFile\immediate\write\auxfile{\noexpand\xdef\noexpand#1{#1}}\fi%
}
\def\section{\Section}
\def\Subsection#1{\global\advance\subsecno by1\relax\medskip %
\leftline{\bf\the\secno.\the\subsecno\ #1}%
\par\nobreak\smallskip\nobreak}
\def\tagsubsection#1{%
\warnIfChanged#1{\the\secno.\the\subsecno}%
\xdef#1{\the\secno.\the\subsecno}%
\ifWritingAuxFile\immediate\write\auxfile{\noexpand\xdef\noexpand#1{#1}}\fi%
}

\def\subsection{\Subsection}

\def\romappno{\uppercase\expandafter{\romannumeral\appno}}
\def\makeNormalizedRomappno{%
\expandafter\makeNormal\expandafter\normalizedromappno%
\expandafter{\romannumeral\appno}%
\edef\normalizedromappno{\uppercase{\normalizedromappno}}}
\def\Appendix#1{\global\advance\appno by1\relax\global\meqno=1\global\secno=0%
\global\subsecno=0%
\bigbreak\bigskip
\centerline{\twelvepoint \bf Appendix %
\romappno. #1}%
\par\nobreak\medskip\nobreak}
\def\tagappendix#1{\makeNormalizedRomappno%
\warnIfChanged#1{\normalizedromappno}%
\xdef#1{\normalizedromappno}%
\ifWritingAuxFile\immediate\write\auxfile{\noexpand\xdef\noexpand#1{#1}}\fi%
}
\def\appendix{\Appendix}
\def\Subappendix#1{\global\advance\subsecno by1\relax\medskip %
\leftline{\bf\romappno.\the\subsecno\ #1}%
\par\nobreak\smallskip\nobreak}
\def\tagsubappendix#1{\makeNormalizedRomappno%
\warnIfChanged#1{\normalizedromappno.\the\subsecno}%
\xdef#1{\normalizedromappno.\the\subsecno}%
\ifWritingAuxFile\immediate\write\auxfile{\noexpand\xdef\noexpand#1{#1}}\fi%
}

\def\eqn#1{\makeNormalizedRomappno%
\ifnum\secno>0%
  \warnIfChanged#1{\the\secno.\the\meqno}%
  \eqno(\the\secno.\the\meqno)\xdef#1{\the\secno.\the\meqno}%
     \global\advance\meqno by1
\else\ifnum\appno>0%
  \warnIfChanged#1{\normalizedromappno.\the\meqno}%
  \eqno({\rm\romappno}.\the\meqno)%
      \xdef#1{\normalizedromappno.\the\meqno}%
     \global\advance\meqno by1
\else%
  \warnIfChanged#1{\the\meqno}%
  \eqno(\the\meqno)\xdef#1{\the\meqno}%
     \global\advance\meqno by1
\fi\fi%
\eqlabeL#1%
\ifWritingAuxFile\immediate\write\auxfile{\noexpand\xdef\noexpand#1{#1}}\fi%
}
\def\defeqn#1{\makeNormalizedRomappno%
\ifnum\secno>0%
  \warnIfChanged#1{\the\secno.\the\meqno}%
  \xdef#1{\the\secno.\the\meqno}%
     \global\advance\meqno by1
\else\ifnum\appno>0%
  \warnIfChanged#1{\normalizedromappno.\the\meqno}%
  \xdef#1{\normalizedromappno.\the\meqno}%
     \global\advance\meqno by1
\else%
  \warnIfChanged#1{\the\meqno}%
  \xdef#1{\the\meqno}%
     \global\advance\meqno by1
\fi\fi%
\eqlabeL#1%
\ifWritingAuxFile\immediate\write\auxfile{\noexpand\xdef\noexpand#1{#1}}\fi%
}
\def\anoneqn{\makeNormalizedRomappno%
\ifnum\secno>0
  \eqno(\the\secno.\the\meqno)%
     \global\advance\meqno by1
\else\ifnum\appno>0
  \eqno({\rm\normalizedromappno}.\the\meqno)%
     \global\advance\meqno by1
\else
  \eqno(\the\meqno)%
     \global\advance\meqno by1
\fi\fi%
}
\def\mfig#1#2{\global\advance\figno by1%
\relax#1\the\figno%
\warnIfChanged#2{\the\figno}%
\edef#2{\the\figno}%
\reflabeL#2%
\ifWritingAuxFile\immediate\write\auxfile{\noexpand\xdef\noexpand#2{#2}}\fi%
}

\def\fig#1{\mfig{fig.\ ~}#1}

\catcode`@=11 

\newif\ifFiguresInText\FiguresInTexttrue
\newif\if@FigureFileCreated
\newwrite\capfile
\newwrite\figfile

\def\PlaceTextFigure#1#2#3#4{%
#3\epsfbox{#4}\hfil\break Figure #1. #2\hfil\break\vskip0.5truein}
\def\PlaceEndFigure#1#2{%
\epsfysize=\vsize\epsfbox{#2}\hfil\break\vfill\centerline{Figure #1.}\eject}

\def\LoadFigure#1#2#3#4{%
\ifundef#1{\phantom{\mfig{}#1}}\fi
\ifFiguresInText
\PlaceTextFigure{#1}{#2}{#3}{#4}%
\else
\if@FigureFileCreated\else%
\immediate\openout\capfile=\jobname.caps%
\immediate\openout\figfile=\jobname.figs%
\fi%
\immediate\write\capfile{\noexpand\item{Figure \noexpand#1.\ }#2.}%
\immediate\write\figfile{\noexpand\PlaceEndFigure\noexpand#1{\noexpand#4}}%
\fi}

\def\listfigs{\ifFiguresInText\else%
\vfill\eject\immediate\closeout\capfile
\immediate\closeout\figfile%
\centerline{{\bf Figures}}\bigskip\frenchspacing%
\input \jobname.caps\vfill\eject\nonfrenchspacing%
\input\jobname.figs\fi}

\font\ninerm=cmr9
\font\eightrm=cmr8
\font\sixrm=cmr6

\def\loadtrueseventeenpoint{
 \font\seventeenrm=cmr10 at 17.28truept
 \font\seventeeni=cmmi10 at 17.28truept
 \font\seventeenbf=cmbx10 at 17.28truept
 \font\seventeenit=cmti10 at 17.28truept
 \font\seventeensl=cmsl10 at 17.28truept
 \font\seventeensy=cmsy10 at 17.28truept
}
\def\loadfourteenpoint{
\font\fourteenrm=cmr10 at 14.4pt
\font\fourteeni=cmmi10 at 14.4pt
\font\fourteenit=cmti10 at 14.4pt
\font\fourteensl=cmsl10 at 14.4pt
\font\fourteensy=cmsy10 at 14.4pt
\font\fourteenbf=cmbx10 at 14.4pt
}
\def\loadtruetwelvepoint{
\font\twelverm=cmr10 at 12truept
\font\twelvei=cmmi10 at 12truept
\font\twelveit=cmti10 at 12truept
\font\twelvesl=cmsl10 at 12truept
\font\twelvesy=cmsy10 at 12truept
\font\twelvebf=cmbx10 at 12truept
}

\font\ninei=cmmi9
\font\eighti=cmmi8
\font\sixi=cmmi6
\skewchar\ninei='177 \skewchar\eighti='177 \skewchar\sixi='177

\font\ninesy=cmsy9
\font\eightsy=cmsy8
\font\sixsy=cmsy6
\skewchar\ninesy='60 \skewchar\eightsy='60 \skewchar\sixsy='60

\font\ninebf=cmbx9
\font\eightbf=cmbx8
\font\sixbf=cmbx6

\font\ninett=cmtt9
\font\eighttt=cmtt8

\hyphenchar\tentt=-1 
\hyphenchar\ninett=-1
\hyphenchar\eighttt=-1

\font\ninesl=cmsl9
\font\eightsl=cmsl8

\font\nineit=cmti9
\font\eightit=cmti8


\newskip\ttglue
\def\tenpoint{\def\rm{\fam0\tenrm}%
  \textfont0=\tenrm \scriptfont0=\sevenrm \scriptscriptfont0=\fiverm
  \textfont1=\teni \scriptfont1=\seveni \scriptscriptfont1=\fivei
  \textfont2=\tensy \scriptfont2=\sevensy \scriptscriptfont2=\fivesy
  \textfont3=\tenex \scriptfont3=\tenex \scriptscriptfont3=\tenex
  \def\it{\fam\itfam\tenit}\textfont\itfam=\tenit
  \def\sl{\fam\slfam\tensl}\textfont\slfam=\tensl
  \def\bf{\fam\bffam\tenbf}\textfont\bffam=\tenbf \scriptfont\bffam=\sevenbf
  \scriptscriptfont\bffam=\fivebf
  \normalbaselineskip=12pt
  \let\sc=\eightrm
  \let\big=\tenbig
  \setbox\strutbox=\hbox{\vrule height8.5pt depth3.5pt width\z@}%
  \normalbaselines\rm}

\def\twelvepoint{\def\rm{\fam0\twelverm}%
  \textfont0=\twelverm \scriptfont0=\ninerm \scriptscriptfont0=\sevenrm
  \textfont1=\twelvei \scriptfont1=\ninei \scriptscriptfont1=\seveni
  \textfont2=\twelvesy \scriptfont2=\ninesy \scriptscriptfont2=\sevensy
  \textfont3=\tenex \scriptfont3=\tenex \scriptscriptfont3=\tenex
  \def\it{\fam\itfam\twelveit}\textfont\itfam=\twelveit
  \def\sl{\fam\slfam\twelvesl}\textfont\slfam=\twelvesl
  \def\bf{\fam\bffam\twelvebf}\textfont\bffam=\twelvebf
                                            \scriptfont\bffam=\ninebf
  \scriptscriptfont\bffam=\sevenbf
  \normalbaselineskip=12pt
  \let\sc=\eightrm
  \let\big=\tenbig
  \setbox\strutbox=\hbox{\vrule height8.5pt depth3.5pt width\z@}%
  \normalbaselines\rm}

\def\fourteenpoint{\def\rm{\fam0\fourteenrm}%
  \textfont0=\fourteenrm \scriptfont0=\tenrm \scriptscriptfont0=\sevenrm
  \textfont1=\fourteeni \scriptfont1=\teni \scriptscriptfont1=\seveni
  \textfont2=\fourteensy \scriptfont2=\tensy \scriptscriptfont2=\sevensy
  \textfont3=\tenex \scriptfont3=\tenex \scriptscriptfont3=\tenex
  \def\it{\fam\itfam\fourteenit}\textfont\itfam=\fourteenit
  \def\sl{\fam\slfam\fourteensl}\textfont\slfam=\fourteensl
  \def\bf{\fam\bffam\fourteenbf}\textfont\bffam=\fourteenbf%
  \scriptfont\bffam=\tenbf
  \scriptscriptfont\bffam=\sevenbf
  \normalbaselineskip=17pt
  \let\sc=\elevenrm
  \let\big=\tenbig
  \setbox\strutbox=\hbox{\vrule height8.5pt depth3.5pt width\z@}%
  \normalbaselines\rm}

\def\seventeenpoint{\def\rm{\fam0\seventeenrm}%
  \textfont0=\seventeenrm \scriptfont0=\fourteenrm \scriptscriptfont0=\tenrm
  \textfont1=\seventeeni \scriptfont1=\fourteeni \scriptscriptfont1=\teni
  \textfont2=\seventeensy \scriptfont2=\fourteensy \scriptscriptfont2=\tensy
  \textfont3=\tenex \scriptfont3=\tenex \scriptscriptfont3=\tenex
  \def\it{\fam\itfam\seventeenit}\textfont\itfam=\seventeenit
  \def\sl{\fam\slfam\seventeensl}\textfont\slfam=\seventeensl
  \def\bf{\fam\bffam\seventeenbf}\textfont\bffam=\seventeenbf%
  \scriptfont\bffam=\fourteenbf
  \scriptscriptfont\bffam=\twelvebf
  \normalbaselineskip=21pt
  \let\sc=\fourteenrm
  \let\big=\tenbig
  \setbox\strutbox=\hbox{\vrule height 12pt depth 6pt width\z@}%
  \normalbaselines\rm}

\def\ninepoint{\def\rm{\fam0\ninerm}%
  \textfont0=\ninerm \scriptfont0=\sixrm \scriptscriptfont0=\fiverm
  \textfont1=\ninei \scriptfont1=\sixi \scriptscriptfont1=\fivei
  \textfont2=\ninesy \scriptfont2=\sixsy \scriptscriptfont2=\fivesy
  \textfont3=\tenex \scriptfont3=\tenex \scriptscriptfont3=\tenex
  \def\it{\fam\itfam\nineit}\textfont\itfam=\nineit
  \def\sl{\fam\slfam\ninesl}\textfont\slfam=\ninesl
  \def\bf{\fam\bffam\ninebf}\textfont\bffam=\ninebf \scriptfont\bffam=\sixbf
  \scriptscriptfont\bffam=\fivebf
  \normalbaselineskip=11pt
  \let\sc=\sevenrm
  \let\big=\ninebig
  \setbox\strutbox=\hbox{\vrule height8pt depth3pt width\z@}%
  \normalbaselines\rm}

\def\eightpoint{\def\rm{\fam0\eightrm}%
  \textfont0=\eightrm \scriptfont0=\sixrm \scriptscriptfont0=\fiverm%
  \textfont1=\eighti \scriptfont1=\sixi \scriptscriptfont1=\fivei%
  \textfont2=\eightsy \scriptfont2=\sixsy \scriptscriptfont2=\fivesy%
  \textfont3=\tenex \scriptfont3=\tenex \scriptscriptfont3=\tenex%
  \def\it{\fam\itfam\eightit}\textfont\itfam=\eightit%
  \def\sl{\fam\slfam\eightsl}\textfont\slfam=\eightsl%
  \def\bf{\fam\bffam\eightbf}\textfont\bffam=\eightbf \scriptfont\bffam=\sixbf%
  \scriptscriptfont\bffam=\fivebf%
  \normalbaselineskip=9pt%
  \let\sc=\sixrm%
  \let\big=\eightbig%
  \setbox\strutbox=\hbox{\vrule height7pt depth2pt width\z@}%
  \normalbaselines\rm}

\def\tenbig#1{{\hbox{$\left#1\vbox to8.5pt{}\right.\n@space$}}}
\def\ninebig#1{{\hbox{$\textfont0=\tenrm\textfont2=\tensy
  \left#1\vbox to7.25pt{}\right.\n@space$}}}
\def\eightbig#1{{\hbox{$\textfont0=\ninerm\textfont2=\ninesy
  \left#1\vbox to6.5pt{}\right.\n@space$}}}

\def\footnote#1{\edef\@sf{\spacefactor\the\spacefactor}#1\@sf
      \insert\footins\bgroup\eightpoint
      \interlinepenalty100 \let\par=\endgraf
        \leftskip=\z@skip \rightskip=\z@skip
        \splittopskip=10pt plus 1pt minus 1pt \floatingpenalty=20000
        \smallskip\item{#1}\bgroup\strut\aftergroup\@foot\let\next}
\skip\footins=12pt plus 2pt minus 4pt 
\dimen\footins=30pc 

\newinsert\margin
\dimen\margin=\maxdimen
\def\titlefont{\seventeenpoint}
\loadtruetwelvepoint 
\loadtrueseventeenpoint

\def\eatOne#1{}
\def\ifundef#1{\expandafter\ifx%
\csname\expandafter\eatOne\string#1\endcsname\relax}
\def\notTrue{\iffalse}\def\isTrue{\iftrue}
\def\ifdef#1{{\ifundef#1%
\aftergroup\notTrue\else\aftergroup\isTrue\fi}}
\def\use#1{\ifundef#1\linemessage{Warning: \string#1 is undefined.}%
{\tt \string#1}\else#1\fi}


\global\newcount\refno \global\refno=1
\newwrite\rfile
\newlinechar=`\^^J
\def\@ref#1#2{\the\refno\n@ref#1{#2}}
\def\n@ref#1#2{\xdef#1{\the\refno}%
\ifnum\refno=1\immediate\openout\rfile=\jobname.refs\fi%
\immediate\write\rfile{\noexpand\item{[\noexpand#1]\ }#2.}%
\global\advance\refno by1}
\def\nref{\n@ref} 
\def\ref{\@ref}   
\def\lref#1#2{\the\refno\xdef#1{\the\refno}%
\ifnum\refno=1\immediate\openout\rfile=\jobname.refs\fi%
\immediate\write\rfile{\noexpand\item{[\noexpand#1]\ }#2\semi}%
\global\advance\refno by1}
\def\cref#1{\immediate\write\rfile{#1\semi}}

\def\preref#1#2{\gdef#1{\@ref#1{#2}}}

\def\semi{;\hfil\noexpand\break}

\def\listrefs{\vfill\eject\immediate\closeout\rfile
\centerline{{\bf References}}\bigskip\frenchspacing%
\input \jobname.refs\vfill\eject\nonfrenchspacing}

\def\inputAuxIfPresent#1{\immediate\openin1=#1
\ifeof1\message{No file \auxfileName; I'll create one.
}\else\closein1\relax\input\auxfileName\fi%
}

\newif\ifWritingAuxFile
\newwrite\auxfile
\def\SetUpAuxFile{%
\xdef\auxfileName{\jobname.aux}%
\inputAuxIfPresent{\auxfileName}%
\WritingAuxFiletrue%
\immediate\openout\auxfile=\auxfileName}


\def\bye{\par\vfill\supereject%
\ifAnyCounterChanged\linemessage{
Some counters have changed.  Re-run tex to fix them up.}\fi%
\end}

\catcode`\@=\active
\catcode`@=12  
\catcode`\"=\active



\def\pol{\varepsilon}

\def\c{\,\cdot\,}

\def\spa#1.#2{\left\langle#1\,#2\right\rangle}
\def\spb#1.#2{\left[#1\,#2\right]}
\def\lor#1.#2{\left(#1\,#2\right)}
\def\sand#1.#2.#3{%
\left\langle\smash{#1}{\vphantom1}^{-}\right|{#2}%
\left|\smash{#3}{\vphantom1}^{-}\right\rangle}
\def\sandp#1.#2.#3{%
\left\langle\smash{#1}{\vphantom1}^{-}\right|{#2}%
\left|\smash{#3}{\vphantom1}^{+}\right\rangle}
\def\sandpp#1.#2.#3{%
\left\langle\smash{#1}{\vphantom1}^{+}\right|{#2}%
\left|\smash{#3}{\vphantom1}^{+}\right\rangle}
\catcode`@=11  
\def\meqalign#1{\,\vcenter{\openup1\jot\m@th
   \ialign{\strut\hfil$\displaystyle{##}$ && $\displaystyle{{}##}$\hfil
             \crcr#1\crcr}}\,}
\catcode`@=12  


\SetUpAuxFile
\loadfourteenpoint
\hfuzz 60 pt

\def\eps{\epsilon}
\def\x#1#2{x_{#1 #2}}
\def\Gbdb{\dot {\overline G}{}_B}
\def\Gbddb{\ddot {\overline G}{}_B}

\def\Gbd{\dot G_B}
\def\Gbdd{\ddot G_B}
\def\sign{\mathop{\rm sign}\nolimits}

\def\dlips{d{\rm LIPS}}

\def\rg{r_\Gamma}

\def\Atree{A^{\rm tree}}
\def\Aloop{A^{\rm 1-loop}}


\def\TABa{1}
\def\TABb{2}
\def\TABc{3}

\def\ref{\nref}
\ref\GravityReview{
B.S. DeWitt, Phys.\ Rev.\ 162:1239 (1967)\semi
M. Veltman, in {\it Les Houches 1975,
Methods in Field Theory}, ed R. Balian and J. Zinn-Justin,
(North Holland, Amsterdam, 1976)}

\ref\StringBased{
Z. Bern and D.A.\ Kosower, Phys.\ Rev.\ Lett.\ 66:1669 (1991);
Nucl.\ Phys.\ B379:451 (1992)}

\ref\PASCOS{Z. Bern and D.A.\ Kosower, in {\it Proceedings of the PASCOS-91
Symposium}, eds.\ P. Nath and S. Reucroft (World Scientific, 1992)}

\ref\FiveGluon{Z. Bern, L. Dixon and D.A. Kosower, Phys.\ Rev. Lett.\
70:2677 (1993)}

\ref\Gravity{Z. Bern, D.C. Dunbar and T. Shimada,
Phys.\ Lett.\ 312B:277 (1993)}

\ref\Zak{M.T. Grisaru and J. Zak, Phys. Lett.\ {90B}:237 (1980)}

\ref\Scherk{ J. Scherk, Nucl.\ Phys. {B31}:222 (1971)\semi
A. Neveu and J. Scherk, Nucl.\ Phys. {B36}:155 (1972)}

\ref\Minahan{
J.\ Minahan, Nucl.\ Phys.\ B298:36 (1988) \semi
K. Miki, Nucl.\ Phys.\ B291:349 (1987) \semi
M.B. Green and N. Seiberg, Nucl.\ Phys.\ B299:559 (1988)}

\ref\GSW{M. B.\ Green, J.H.\ Schwarz,
and E.\ Witten, {\it Superstring Theory} (Cambridge University
Press) (1987)}

\ref\KLT{H. Kawai, D.C. Lewellen and  S.H.H. Tye,
Nucl.\ Phys.\ B269:1 (1986)}

\ref\Berends{F.A. Berends, W.T.\ Giele and H. Kuijf,
Phys. Lett.\ {211B}:91 (1988)}

\ref\Tasi{
Z. Bern, hep-ph/9304249, in {\it Proceedings of Theoretical
Advanced Study Institute in High Energy Physics (TASI 92)},
eds.\ J. Harvey and J. Polchinski (World Scientific, 1993)}

\ref\Weak{
Z.\ Bern and A.\ Morgan, Phys.\ Rev.\ D49:6155 (1994)}

\ref\Background{G. 't Hooft,
Acta Universitatis Wratislavensis no.\
38, 12th Winter School of Theoretical Physics in Karpacz; {\it
Functional and Probabilistic Methods in Quantum Field Theory},
Vol. 1 (1975)\semi
B.S.\ DeWitt, in {\it Quantum Gravity II}, eds. C. Isham, R.\ Penrose and
D.\ Sciama (Oxford, 1981)\semi
L.F.\ Abbott, Nucl.\ Phys.\ B185:189 (1981)\semi
L.F\ Abbott, M.T.\ Grisaru and R.K.\ Schaeffer,
Nucl.\ Phys. {B229}:372 (1983)}

\ref\SuperSpace{M.T. Grisaru and  W. Siegel, Nucl.\ Phys.\ B187:149 (1981)}

\ref\GGRS{S.J. Gates, M.T. Grisaru, M. Rocek and W. Siegel,
 {\it Superspace}, (Benjamin/Cummings, 1983)}

\ref\Mapping{Z. Bern and D.C.\ Dunbar,  Nucl.\ Phys.\ B379:562 (1992)}

\ref\FirstQ{
E.S.\ Fradkin and A.A.\ Tseytlin, Phys. Lett. 158B:316 (1985);
163B:123 (1985); Nucl. Phys. B261:1 (1985)\semi
M. Strassler, Nucl.\ Phys.\ {B385}:145 (1992)\semi
M.G. Schmidt and C. Schubert Phys.\ Lett.\ 318B:438 (1993)\semi
D.G.C.\ McKeon, Ann. Phys. (N.Y.) 224:139 (1993)}

\ref\Green{
M.B.\ Green, J.H.\ Schwarz and L. Brink, Nucl.\ Phys.\ B198:472 (1982)}

\ref\Cutkosky{L.D.\ Landau, Nucl.\ Phys.\ 13:181 (1959)\semi
 S. Mandelstam, Phys.\ Rev.\ 112:1344 (1958), 115:1741 (1959)\semi
 R.E.\ Cutkosky, J.\ Math.\ Phys.\ 1:429 (1960)}

\ref\SusyFour{Z. Bern, D.C. Dunbar, L. Dixon and D.A. Kosower,
hep-ph/9403226 UCLA/94/4, to appear in Nucl.\ Phys.\ B}

\ref\SusyOne{Z. Bern, D.C. Dunbar, L. Dixon and D.A. Kosower,
hep-ph/9405248 UCLA/94/17, and UCLA/94/29 in preparation }

\ref\HVb{G. 't\ Hooft and M.\ Veltman,
Ann. Inst. Henri Poincar\'e 20:69 (1974)}

\ref\Matter{S. Deser and  P. van Nieuwenhuizen,
Phys.\ Rev.\ D10:411 (1974)\semi
M.T. Grisaru, P. van Nieuwenhuizen and  C.C.\ Wu,
Phys.\ Rev.\ D12:1813 (1975)}

\ref\GNV{ M.T.\ Grisaru, P.\  van Nieuwenhuizen and J.A.M. Vermaseren,
Phys.\ Rev.\ Lett.\ 37:1662 (1976)}

\ref\Bosonic{ Z. Bern, Phys.\ Lett.\ 296B:85 (1992)}

\ref\SusyReg{W. Siegel, Phys.\ Lett.\ 84B:193 (1979)\semi
D.M.\ Capper, D.R.T.\ Jones and P. van Nieuwenhuizen, Nucl.\ Phys.\
B167:479 (1980)\semi
L.V.\ Avdeev and A.A.\ Vladimirov, Nucl.\ Phys.\ B219:262 (1983)\semi
I.\ Jack, D.R.T.\ Jones and K.L. Roberts,  hep-ph/9401349}

\ref\SpinorGravity{
S.F.Novaes and D.Spehler,
Phys.Rev. D44:3990 (1991); Nucl.Phys.\ B371:618(1992)\semi
H.T.\ Cho, K.L.\ Ng, Phys.\ Rev.\ D47:1692 (1993) }

\ref\XZC{%
F.\ A.\ Berends, R.\ Kleiss, P.\ De Causmaecker, R.\ Gastmans and T.\ T.\ Wu,
        Phys.\ Lett.\ 103B:124 (1981)\semi
P.\ De Causmaeker, R.\ Gastmans,  W.\ Troost and  T.\ T.\ Wu,
Nucl. Phys. B206:53 (1982)\semi
R.\ Kleiss and W.\ J.\ Stirling,
   Nucl.\ Phys.\ B262:235 (1985)\semi
   J.\ F.\ Gunion and Z.\ Kunszt, Phys.\ Lett.\ 161B:333 (1985)\semi
 R.\ Gastmans and T.T.\ Wu,
{\it The Ubiquitous Photon: Helicity Method for QED and QCD} (Clarendon Press)
(1990)\semi
Z.\ Xu, D.-H.\ Zhang and L. Chang, Nucl.\ Phys.\ B291:392 (1987)}

\ref\Susy{ M.T. Grisaru, H.N. Pendleton and P.  van Nieuwenhuizen,
Phys. Rev. {D15}:996 (1977)\semi
M.L.\ Mangano and S.J. Parke, Phys.\ Rep.\ {200}:301 (1991)}

\ref\Sannan{S.\ Sannan, Phys.\ Rev.\ D34:1748 (1986)}

\ref\Flip{ Y.Q. Cai and  G. Papini, Phys.\ Rev.\ Lett.66:1259,(1991)\semi
R. Aldrovandi,G.E.A. Matsas, S.F. Novaes and
D. Spehler, gr-qc/9404018}

\ref\NieWu{P. van Niewenhuizen and C.C.\ Wu, J. Math. Phys. 18:81 (1977)}

\ref\GoroffSagnotti{ M.H. Goroff and A. Sagnotti,
Phys. Lett. 160B:81(1985), Nucl. Phys. B266:709 (1986)}

\ref\VanDVen{A.E.M.\ van de Ven,  Nucl.\ Phys.\ B378:309 (1992)}

\ref\Kaj{K. Roland, Phys.\ Lett.\ 289B:148 (1992) \semi
G. Cristofano, R. Marotta and K. Roland, Nucl.\ Phys.\ B 392:345 (1993)}

\ref\Wline{M.G. Schmidt and C. Schubert, hep-th/9403158 \semi
C.S.\ Lam, hep-ph/9406388}

\nopagenumbers

\noindent

$\null$

\vskip -1.6 cm

hep-th/9408014

\rightline{UCLA/TEP/94/30}
\rightline{SWAT-94-37}

\hfill July 1994


\vskip 1in
{\titlefont\centerline{\bf Calculation of Graviton Scattering Amplitudes}}
{\titlefont\centerline{\bf using String-Based Methods}}
\vskip .5in
\vskip .3 cm


\centerline{\bf David C. Dunbar${}^{\dagger}$ }
\centerline{\it Department of Physics}
\centerline{\it UCLA}
\centerline{\it Los Angeles, CA 90024}
\centerline{\it USA }

\smallskip \centerline{\rm and} \smallskip

\vglue 0.2cm
\centerline{\bf Paul S. Norridge}
\centerline{\it Department of Physics}
\centerline{\it University College of Swansea}
\centerline{\it Swansea SA2 8PP}
\centerline{\it UK}

\vskip 1.2 truecm \baselineskip12pt


\centerline{\bf Abstract }

\vskip 0.1 truecm
{
\narrower\smallskip
\smallskip
Techniques based upon the string organisation of amplitudes
may be used to simplify field theory calculations.
We apply these techniques to perturbative gravity
and calculate all one-loop amplitudes for four-graviton scattering
with arbitrary internal particle content.
Decomposing the amplitudes into contributions arising from
supersymmetric multiplets greatly simplifies these calculations.
We also discuss how unitarity may be used to constrain
the amplitudes.
}

\baselineskip14pt

\vglue 0.3cm

\vfil\vskip .2 cm
\noindent\hrule width 3.6in\hfil\break
${}^{\dagger}$Address after Sept. 1, 1994: University
College of Swansea,
UK. \hfil\break

\vfill\eject

\footline={\hss\tenrm\folio\hss}

\section{Introduction}

Calculations in perturbative gravity are well known to be
prohibitively difficult using conventional Feynman diagram techniques
[\use\GravityReview]. In typical gauges, the graviton vertices contain
considerably more terms than in gauge theories
and two powers of loop momentum
rather than one. These two features help to make perturbative gravity
calculation an algebraic nightmare.

Recently, the alternate organisation of amplitudes
offered by string theory has been used by Bern and Kosower
to construct rules for
computations in gauge theories [\use\StringBased,\use\PASCOS].
These rules allow a considerable algebraic simplification compared to
normal Feynman diagram techniques as evidenced by the first calculation
of the five-gluon one loop amplitudes [\use\FiveGluon].
Since string theory also contains gravity
one can apply these
techniques to obtain rules for calculations in perturbative gravity.
In ref.~[\use\Gravity], the string-based technique for perturbative
gravity was outlined and a sample calculation of the
$A^{\rm one-loop}(-,+,+,+)$
four graviton helicity amplitude was performed.
This is the simplest of the four graviton amplitudes being finite and
without cuts and has been previously
calculated by first calculating the contributions to the four graviton
amplitude from real scalars in the loop and then using
the supersymmetry Ward identities
[\use\Zak].

In this paper we present a detailed description of the rules for
one-loop $n$-graviton amplitudes  and
use these
to calculate
the four graviton amplitude for
all helicity
configurations with arbitrary particle content in the loop.
As with the QCD method,
these rules arise by looking at the infinite tension limit
[\use\Scherk,\use\Minahan] of a string theory.
However, these rules can be used with no knowledge of
string theory.
The rules are in many ways a ``double'' copy of those for QCD,
$$
(\hbox{Gravity}) \sim (\hbox{Yang-Mills})^2 \; .
\eqn\GravityYM
$$
which reflects
the fact that in string theory, a closed string may be regarded as the
product of two open strings [\use\GSW]
$$
(\hbox{Closed String}) \sim (\hbox{Open String})^2 \; .
\anoneqn
$$
This equivalence is largely true at the level of the integrands of
diagrams.  (There is, however, a small amount of interference between
the two ``halves'' of the string which is related to the zero mode
integral in string theory.)  Using tree-level relationships which
embody (\use\GravityYM) [\use\KLT] string theory has been used to
calculate tree-level graviton amplitudes previously in
ref.~[\use\Berends].

As inspired by the string-based method, we implement a supersymmetric
decomposition of the amplitudes similar to that recently used for
gauge theories [\use\Tasi,\use\Weak].  Instead of calculating the
contributions to the loop amplitudes from individual particles
circulating (which may be the graviton, gravitino, vector, Weyl
fermion or scalar in a gravitational theory) we calculate the
contributions from various supersymmetric multiplets plus the scalar
contribution. Corresponding to the five particle types we must
calculate four supersymmetric contributions (which we choose to be the
$N=1$, $N=4$, $N=6$ and $N=8$ matter multiplets) plus the scalar. The
contribution from any individual particle type is just a linear
combination of these.  This decomposition enables us to exploit the
simplifications found in supersymmetry calculations.  In
supersymmetric theories there are cancellations between the fermions
and bosons.  If one uses a suitable formalism, these cancellations are
manifest diagram by diagram. This proves to be an enormous
simplification.  Examples of such beautiful formalisms are 1) the
string based rules and 2) a superspace formalism using the background
field method [\use\Background,\use\SuperSpace,\use\GGRS].  In a
general gauge even in a superspace formalism the cancellations do not
occur diagram by diagram.  The relationship between string based rules
for gauge theories and conventional field theory is particularly close
when the field theory is organised using the background field method
[\use\Mapping,\use\FirstQ].

For the four-point function, without cancellations, the Feynman
parameter integral generically has eight powers of Feynman parameters
(or equivalently eight powers of loop momentum).  For the $N=1$ matter
multiplet (containing a scalar and a fermion) there is a cancellation
of the two leading powers of Feynman parameters which simplifies the
calculation considerably. With increasing $N$ more cancellations occur
until for the maximal case, $N=8$, all eight powers cancel and one is
left with a trivial sum of scalar box integrals.  The special cases of
$N=8$ supergravity and $N=4$ super-Yang-Mills four point functions
were obtained previously in ref.~[\use\Green] also be examining the
infinite string tension limit.

Another advantage of the supersymmetric decomposition is that the
supersymmetric amplitudes, again due to the cancellations in loop
momentum, can be strongly constrained by unitarity via the Cutkosky
rules [\use\Cutkosky].  In ref.~[\use\SusyFour,\use\SusyOne],
situations in a gauge theory where unitarity completely determines the
one-loop amplitudes are given.  For the four-point gravity calculation
the $N=8$ and $N=6$ contributions can be determined completely using
the Cutkosky rules and are in agreement with the string-based results.
The Cutkosky rules are also used to check the cuts in the remaining
amplitudes. As a further example of the uses of the Cutkosky rules we
calculate the logarithmic part of the two-loop amplitude
$A^{\rm 2-loop}(++++)$.

As is well known, pure gravity is renormalisable at one-loop
[\use\HVb] whereas gravity coupled to matter is not [\use\Matter] .
However, the ultra-violet infinities do not arise in amplitudes
containing only external gravitons but appear in amplitudes with
external matter [\use\GNV].  Our amplitudes are all ultra-violet
finite in agreement with the formal arguments. The amplitude with
gravitons circulating in the loop contain the infra-red singularities
as expected.

\section{Rules for one-loop gravity}

In ref.~[\use\StringBased,\use\PASCOS] rules were introduced for the
calculation
of gauge theory amplitudes. These were obtained by taking the infinite
string tension limit of string theory amplitudes and can be used
instead of Feynman diagram techniques.
Although derived from string theory, they can be used without explicit
knowledge of string theory. The various contributions to the amplitude
are associated with $\phi^3$ diagrams. Typically, the string organisation
lead to a more compact integrand for these diagrams than that arising
in conventional field theory. The string-based technique has been used
to perform significant  calculations such as the five-gluon one-loop
contributions [\use\FiveGluon].

There are two slightly different but equivalent formulations of the
rules; One is obtained by taking the infinite tension limit of a
superstring [\use\StringBased,\PASCOS] whereas the other is obtained
by taking the infinite tension limit of a bosonic string
[\use\Bosonic,\Tasi].  The bosonic form of the rules is more compact
but the appropriate rules for fermions circulating in the loop must be
inferred from the superstring case.  The string-based rules for
one-loop gravity which we present here were outlined in
ref.~[\use\Gravity]. These are based upon the bosonic formulation. The
rules for gravity have many similarities to those for gauge theories
so we will be brief in presenting them.

The initial step in the rules is to draw all $\phi^3$ diagrams,
excluding tadpoles.  There is also no need to include diagrams with a
loop isolated on an external leg since these vanish when dimensional
regularisation is used.  The external legs of these diagrams should be
labeled, with diagrams containing all orderings included.  The inner
lines of trees attached to the loop are labeled according to the rule
that as one moves form the outer lines to the inner ones, one labels
the inner line with the same label as the most clockwise of the two
outer lines attached to it.  The contribution from each labeled
$n$-point $\phi^3$-like diagram with $n_{\ell}$ legs attached to the
loop is
$$
\eqalign{
{\cal D} =
i { (-\kappa)^n \over (4\pi)^{2-\eps} }
 \Gamma(n_\ell-2+\eps)
&\int_0^1 dx_{i_{n_\ell-1}} \int_0^{x_{i_{n_\ell-1}}} dx_{i_{n_\ell-2}} \cdots
\int_0^{x_{i_3}}  dx_{i_2} \int_0^{x_{i_2}} dx_{i_1} \cr
& \times
{K^{}_{\rm red}(x_{i_1},\dots,x_{i_{n_\ell}}) \over
\Bigl(\sum_{l<m}^{n_\ell} P_{i_l}\c P_{i_m} \x {i_m}{i_l}
(1-\x {i_m}{i_l})\Bigr)^{n_\ell -2  +\eps}}  \cr}
\eqn\IntegrationRule
$$
where the ordering of the loop parameter integrals corresponds to the
ordering of the $n_\ell$ lines attached to the loop,
$x_{ij} \equiv x_i - x_j$. The $x_{i_m}$
are related to ordinary Feynman parameters
by $x_{i_m} = \sum_{j=1}^m a_j$. $K_{\rm red}$ is the ``reduced
kinematic factor'', which the string-based rules efficiently yield
in a compact form.  The lines attached to the loop carry
momenta $P_i$ which will be off-shell if there is a tree attached to
that line.  The dimensional regularisation parameter $2\eps = 4 - D$
handles all ultra-violet and infra-red divergences.  The amplitude is
then given by summing over all diagrams.
We also use the equivalent
Schwinger proper-time form of the amplitude
$$
\eqalign{
{\cal D}_S
&=
i { (-\kappa)^n \over (4\pi)^{2-\eps} }
\int \prod_{i=1}^{n_\ell-1} dx_i \prod_{i<j}^{n_\ell}
 \int_0^\infty dT\;  T^{n_{\ell}-3+\eps}
\exp\Bigl( -T \sum_{l<m}^{n_\ell} P_{i_l}\c P_{i_m} \x {i_m}{i_l}
(1-\x {i_m}{i_l}) \Bigr) \cr
& \hskip 5.0 truecm
\times {K^{}_{\rm red}(x_{i_1},\dots,x_{i_{n_\ell} },T) }
\cr}
\eqn\DaveKine
$$
as discussed in ref.[\use\Mapping]

In order to evaluate $K_{\rm red}$, one starts with the graviton
kinematic expression
$$
\eqalign{
{\cal K} &=
\int \prod_{i=1}^n dx_i d \bar x_i \prod_{i<j}^n
\exp\biggl[ k_i\c k_j G_B^{ij} \biggr]
\exp \biggl[ (k_i\c\pol_j - k_j\c\pol_i) \, \Gbd^{ij}
         - \pol_i\c\pol_j\, \Gbdd^{ij} \biggr] \cr
& \hskip 1 cm \times
\exp \biggl[ (k_i\c\bar\pol_j - k_j\c\bar\pol_i) \, \Gbdb^{ij}
         - \bar\pol_i\c\bar\pol_j\, \Gbddb^{ij} \biggr]
\exp \biggl[- ( \pol_i\c\bar\pol_j + \pol_j\c\bar\pol_i )
\,  H_B^{ij} \biggr]
\biggr|_{\rm multi-linear} \cr}
\eqn\MasterKin
$$
where the `multi-linear' indicates that only the terms linear in all
$\pol_i$ and $\bar\pol_i$ are included.  The graviton polarization
tensor is reconstructed by taking
$\pol_i^\mu\bar\pol_i^\nu \rightarrow \pol_i^{\mu\nu}$.
{}From a string theory perspective $G_B$ is the bosonic Green function
on the string world sheet, $\Gbd$ and $\Gbdd$ are derivatives of this
Green function with respect to left-moving variables, and $\Gbdb$ and
$\Gbddb$ are derivatives with respect to right-movers.
(Since a closed string is periodic the variables described the string
world sheet can split into ``left-moving'' and ``right-moving''.)
The
term $H_B^{ij}$ is the derivative of the Green function with
respect to one left mover and one right mover variable.
The functions $G_B^{ij}$,$\Gbdd^{ij}$ and $H_B^{ij}$ are to
taken as symmetric in the $i$ and $j$ indices while
$\Gbd$ is antisymmetric.
Although the above expression contains much information in string
theory, when one takes the
infinite string tension limit [\use\StringBased,\PASCOS] it
should
merely be regarded as a function
which contains all the information
necessary to generate $K_{red}$ for all graphs.
The utility of the string based method
partially lies in this compact
representation (which is valid for arbitrary numbers of legs!).
The existence of an
overall function which reduces to the Feynman parameter polynomial for each
diagram is one of the most useful features of the string based rules.

The appropriate expression for gauge theories is obtained
by setting $\bar\pol=0$ in the above. The gravity expression is like
a double copy of the gauge theory expression apart from the
$H_B^{ij}$ terms which mix the left and right movers.

The first step in applying the rules is to remove all of the
$\Gbddb^{ij}$ and $\Gbdd^{ij}$ by integrating the kinematic expression
by parts with respect to the variables $x_i$
and $\bar x_i$ where necessary.
When manipulating
this formula we take
$\Gbd^{ij}$, $\Gbdd^{ij}$, $\Gbdb^{ij}$ and $\Gbddb^{ij}$
to
mean $\partial_{x_i} G_B^{ij}$, $\partial_{x_i}^2 G_B^{ij}$,
$\partial_{\bar x_i} G_B^{ij}$ and
$\partial_{\bar x_i}^2 G_B^{ij}$ respectively.
(After direct substitution of the values of the functions
in the field theory limit these relations are almost but not quite true;
however, for the purposes of
manipulating eq.~(\use\MasterKin)
this distinction
is unimportant.)
While carrying out this process one must
take into account the cross-terms where a left-mover derivative hits
right-mover terms, and vice versa.  This can be done by using the
results
$$
\eqalign{
{ \partial\over \partial{\bar x_k} }
{\Gbd^{ij} } =
{\delta_{ki}H_B^{ij}-\delta_{kj}H_B^{ij} }
& \hskip 1 cm
{ \partial\over \partial{x_k} } {\Gbdb^{ij} }
=
{\delta_{ki}H_B^{ij}-\delta_{kj}H_B^{ij} }
 \cr
{\partial\over \partial{\bar x_k} } \Gbdd^{ij} = 0
\ \ \ \ \ \ \ \ \ \ \  & \hskip 1 cm
{ \partial\over \partial{ x_k} }\Gbddb^{ij} = 0}
$$
For example,  if the expression
$$
\eqalign{
&
\int \prod_{i=1}^4 dx_i d \bar x_i \prod_{i<j}^4
\exp\biggl[ k_i\c k_j G_B^{ij} \biggr]
\Gbd^{34}
\Gbdd^{12}\Gbdb^{13}(\Gbdb^{34})^2
}
$$
is integrated by parts with respect to $x_1$, the result is
$$
\eqalign{
\int \prod_{i=1}^4 dx_i d \bar x_i \prod_{i<j}^4
\exp\biggl[ k_i\c k_j G_B^{ij} \biggr]&\Gbd^{34}
\Gbd^{12} (\Gbdb^{34})^2\cr
&\times\biggl( \bigl( k_1\c k_2 \Gbd^{12} +
 k_1\c k_3 \Gbd^{13} + k_1\c k_4 \Gbd^{14} \bigr)
\Gbdb^{13}
+H_B^{13}
\biggr).}
$$

Having carried out the integration by parts
we now may carry out simple substitution rules for each
diagram to obtain $K_{red}$.
First
the $ \prod_{i<j}^n\exp\bigl[ k_i\c k_j
G_B^{ij} \bigr]$ term and the integrals over
$x_i$ and $\bar x_i$ are dropped
from the kinematic expression. (Since the appropriate
contributions have been included in the rules).
After integration by parts, ${\cal K}$ will be a
sum of terms each of which has $n$ $\Gbd$ and $n$ $\Gbdb$.
(An $H_B$ is equivalent to one $\Gbd$ and one $\Gbdb$.)

Any diagram will be a loop with $n_{\ell}$ legs attached with possible
non-trivial trees attached to the loop. The rules have two parts.
Firstly tree rules are applied to ${\cal K}$. These produce a
truncated ${\cal K}$ which is a series of terms each with $n_{\ell}$
$\Gbd$ and $\Gbdb$. Secondly loop substitution rules are applied which
give the Feynman parameter polynomial for the diagram.  The tree rules
are applied iteratively working from the outside of the attached trees
towards the loop. For a two-point tree with outer legs labeled by $i$
and $j$, one carries out the substitutions
$$
\eqalign{
&(\Gbd^{ij})^n (\Gbdb^{ij})^m
\rightarrow  \delta_{n,1} \delta_{m,1}
{ 1\over (-2 k_i \c k_j) }  \cr
&i \rightarrow j \hskip 2 cm \hbox{in remaining factors} }
\anoneqn
$$
in each term.  This should be applied at each tree vertex.

Once the tree rules have been carried out for a diagram, one applies
the loop rules.  These depend on the particles circulating in the
loop.  They are essentially independent applications of the Yang-Mills
rules to the left- and right-mover parts, with an extra substitution
for cross-terms $H_B$.

One-loop amplitudes depend upon the particle circulating in the loop.
and the substitution rules are corresponding different
for different particle types.
For gauge theories
there are three types of particles/rules, those for scalars $S$,
fermions $F$ and vectors $V$. For gravity, the rules are a double
copy of the gauge theory rules and this choice of substitutions can be
chosen differently for the two copies.
That is, we can apply different substitution rules to the
$\Gbd$ (left-movers) and the $\Gbdb$ (right movers).
The particle content circulating in the loop corresponding to
these choices of the loop substitution rules is given in table~\use\TABa .

\vskip .5 cm
\hskip 1.9 truecm
\hbox{
\def\tend{\cr \noalign{\hrule}}

\vbox{\offinterlineskip
{
\hrule
\halign{
        &\vrule#
        &\strut\quad\hfil #\hfil\quad\vrule
        & \quad\hfil\strut # \hfil 
        \cr
height13pt  &{\bf Substitution}  &{\bf Particle Content}      &\tend
height12pt  & $2[S,S]$
& complex scalar  &\tend
height12pt  & $-2[S,F]$
& Weyl Fermion &\tend
height12pt  & $2[S,V]$
& Vector &\tend
height12pt   &  $-4[V,F]$
& gravitino and Weyl Fermion &\tend
height12pt   &  $4[V,V]$
& graviton and complex scalar  &\tend
height12pt   &  $4[V,V]-2[S,S]$
& graviton   &\tend
height12pt   &  $-4[V,F]+2[S,F]$
& gravitino   &\tend
}
}
}
}
\nobreak
{\baselineskip 10 pt
\narrower\smallskip\noindent\ninerm
{\ninebf Table~\TABa :} Applying the substitution rules shown corresponds
to having the particle content shown circulating in the loop.
$[x,y]$ denotes applying substitution rules
$x$ and $y$ to $\Gbd$ and $\Gbdb$.
\smallskip}

$F$ and $V$ each produce two types of contribution.  The
first contribution is just the scalar $S$
but the second is different in the
two cases. The different contribution we refer to as
the ``cycle''  contribution $C_V$ and $C_F$.
$$
\eqalign{
F=& S +C_F \cr
V=& S +C_V
\cr}
\anoneqn
$$
The common $S$ contribution is obtained by
making the substitutions
$$
\eqalign{
\Gbd^{ij} &\longrightarrow  {1\over 2} (-\sign(\x ij ) + 2 \x ij )
\cr
\Gbdb^{ij} &\longrightarrow  {1\over 2} (-\sign(\x ij ) + 2 \x ij )
\cr
H_B^{ij}  &\longrightarrow  {1\over 2 T} \cr}
\eqn\Gsubs
$$
in the Schwinger parameterization (\use\DaveKine).
(Before taking the infinite tension limit a $\delta$-function
exist in $H_B^{ij}$ however as discussed in ref.~[\use\Minahan]
this $\delta$-function does not contribute in the infinite string
tension limit of physical amplitudes.)
The cycle contribution comes
from ``cycles'' of $\Gbd$. A cycle is  a sequence of $\Gbd$'s
$$
\Gbd^{i_1i_2}\Gbd^{i_2i_3}\dots\Gbd^{i_mi_1}
$$
The substitution rules for these cycles is different in the three
cases. For the scalar they are vanishing.
For $C_V$, the  substitution rules are
$$
\eqalign{
\Gbd^{i_1 i_2}\Gbd^{i_2 i_1} &\rightarrow 1 \cr
\Gbd^{i_1 i_2}\Gbd^{i_2 i_3}\dots
\Gbd^{i_{m-1} i_m}\Gbd^{i_mi_1} &\rightarrow 1/2 \hskip 1 cm (m>2)}
$$
where all the cycles must follow the ordering of the legs, and only
one cycle at a time may contribute to any term.  Once these
substitutions have been made all remaining $\Gbd$'s should be replaced
as in eq.~(\use\Gsubs).
For $C_F$ the following substitution is made
$$
\Gbd^{i_1i_2}\Gbd^{i_2 i_3}\dots
\Gbd^{i_{m-1}  i_m}\Gbd^{i_m i_1} \rightarrow 
- (-1/2)^m \prod^m_{k=1}sign(x_{{i_k}{i_{k+1}}})
$$
In contrast to the $V$ rules, all cycles contribute in the $F$
case regardless of ordering.  Also, all combinations of one or more
cycles from each term contribute. Again, once these substitutions have
been made all remaining $\Gbd$'s should be replaced as in (\Gsubs).

For example if, for the four-point amplitude, we have a term in the
Kinematic expression
$$
K=(\Gbd^{12})^2(\Gbd^{34})^2
\anoneqn
$$
Then for the box diagram with ordering of legs $1234$ the cycle
contributions for the two cases are
$$
\eqalign{
C_F: \; K &\longrightarrow -{1\over 4}\Bigl( {1\over 2}(1+2 \x 34) \Bigr)^2
                       -{1\over 4}\Bigl( {1\over 2}(1+2 \x 12) \Bigr)^2
                       +{1\over 16}
\cr
C_V: \; K &\longrightarrow \Bigl({1\over 2}(1+2 \x 34) \Bigr)^2
                       +\Bigl({1\over 2}(1+2 \x 12) \Bigr)^2
\cr}
\eqn\CycleCases
$$
there being no cycle contribution in the scalar case.

This process gives an expression for $K_{\rm red}$ for each diagram
for arbitrary particle content in the loop.  The integral in
(\IntegrationRule) can now be carried out.  The contributions from
each diagram are then summed over. Explicit simple examples of the
applications of the string based rules for QCD are given in
refs.~[\PASCOS,\Tasi] and for gravity in ref.~[\Gravity] which the
interested reader may wish to examine to see the simplicity of the
string based method.  The string based rules have advantages in
producing compact expressions for the numerators in the Feynman
parameter integrals.

\section{Supersymmetric Decomposition}

A useful way to organise the $n$-graviton amplitudes is to use a
``supersymmetric decomposition''. The loop amplitudes depends upon the
state circulating in the loop.  In a graviton scattering calculation
this may be one of five states: a scalar, a Weyl fermion, a vector, a
gravitino or the graviton itself. The string based rules can be used
to calculate these contributions individually. However, it proves
convenient to calculate the contributions from supersymmetric
multiplets instead.  Amplitudes for all choices of particles in the
loop can be written as linear combinations of those for certain
choices of supersymmetric multiplets and for a scalar. In particular
we choose one multiplet from each of N=1,4,6,8 supersymmetric theories
with particle content given in the following table. These multiplets
are centered around the spin-0 complex scalar.

\vskip .5 cm
\centerline{
\hbox{
\def\tend{\cr \noalign{\hrule}}

\vbox{\offinterlineskip
{
\hrule
\halign{
        &\vrule#
        &\strut\quad\hfil #\hfil\quad\vrule
        &\strut\quad\hfil #\hfil\quad\vrule
        &\strut\quad\hfil #\hfil\quad\vrule
        &\strut\quad\hfil #\hfil\quad\vrule
        &\strut\quad\hfil #\hfil\quad\vrule
        & \quad\hfil\strut # \hfil 
        \cr
height13pt  &{\bf $N$ }  & scalars & spin-1/2  & spin-1 & spin-3/2  & spin-2
&\tend
height12pt  &  $N=0$
&  1 &   &  & &  &\tend
height12pt  &  $N=1$
&  1 & 1  &  & &  &\tend
height12pt  & $N=4$
&  3  & 4   & 1 & &  &\tend
height12pt  & $N=6$
&  10 & 15 & 6 & 1 &  &\tend
height12pt   &  $N=8$
& 35 & 56 & 28 & 8 & 1
&\tend
}
}
}
}}
\nobreak
{\baselineskip 10 pt
\narrower\smallskip\noindent\ninerm
{\ninebf Table~\TABb :}
Particle content of the supersymmetric multiplets we consider.
Scalars are complex, and the fermions are Weyl.
\smallskip}

This decomposition is useful when
evaluating amplitudes. In general, the integrations involved in
amplitudes increases considerable with the degree of the
polynomial in the numerator of the
Feynman parameter
integral. (Or equivalently with the degree of the loop
momentum polynomial if performing momentum integrals.)
If one uses the string-based rules cancellations due to
supersymmetry occur within each diagram,
reducing the complexity of computations. Similar cancellations
occur if one uses a background field method within a
superfield formalism [\use\GGRS]. (The relationship between
background field methods and string based calculations is
explored in [\use\Mapping].)

Specifically, within the string based rules these simplifications can
be seen as cancellations between the common contributions within
multiplets.  In terms of Feynman parameters, for a general $n$-point
integral the scalar term $S$ is a polynomial of degree $n$ however the
cycle contributions are polynomials of degree $n-2$.  For each
particle type, there is a scalar contribution $N_s [S,S]$ where $N_s$
counts the degrees of freedom with fermions having negative
weight. Hence for any supersymmetric multiplet the $[S,S]$ term will
cancel and the Feynman parameter polynomial will be simplified. With
increasing $N$ there are increasing cancellations. Also for the
combination $C_V-4C_F$ the two- and three-cycle contributions cancel
leaving a polynomial of degree $n-4$.  This may be seen, for example,
by comparing the cycle contributions in eqn.~(\CycleCases).  For the
supergravity multiplets cancellations can occur on both left and right
movers.  For example, for the $N=8$ calculation we have
$$
A^{N=8}=A^{\rm graviton}
-8A^{\rm gravitino}+28A^{\rm vector}-56A^{\rm
fermion}+30A^{\rm scalar}
\anoneqn
$$
Inserting the rules from
table~\use\TABa\ we find
$$
A^{N=8}=4[C_V,C_V]-32[C_V,C_F]+64[C_F,C_F]
=4[C_V-4C_F,C_V-4C_F]
\anoneqn
$$
{}From this we see that for a $n$-point
integral the Feynman parameter polynomial would be $2n-8$ at most. The
cancellations for a given $N$ are shown in table~3.
(We are using a regularisation scheme which preserves supersymmetry [\SusyReg]
which simplifies the decomposition. In other regularisation schemes the form
is the decomposition is a little more complex.)

\vskip .5 cm
\hskip 1.9 truecm
\hbox{
\def\tend{\cr \noalign{\hrule}}

\vbox{\offinterlineskip
{
\hrule
\halign{
        &\vrule#
        &\strut\quad\hfil #\hfil\quad\vrule
        &\strut\quad\hfil #\hfil\quad\vrule
        & \quad\hfil\strut # \hfil 
        \cr
height13pt  &{\bf $N$ }  & Contribution  & Degree
&\tend
height12pt  &  $N=0$
& $2[S,S]$ & $2n$ &\tend
height12pt  &  $N=1$
& $2[C_F,S]$ & $2n-2$    &\tend
height12pt  & $N=4$
& $2[C_V-4C_F,S]$  & $2n-4$      &\tend
height12pt  & $N=6$
& $-4[C_V-4C_F,C_F]$  & $2n-6$    &\tend
height12pt   &  $N=8$
& $4[C_V-4C_F,C_V-4C_F]$  & $2n-8$
&\tend
}
}
}
}
\nobreak
{\baselineskip 10 pt
\narrower\smallskip\noindent\ninerm
{\ninebf Table~\TABc :}
{The String rules appropriate for the multiplet
are given and the degree of the Feynman parameter polynomial for
an $n$-point loop integral.}
}

\vskip 0.2 cm

To reconstruct the amplitudes for
specific particles in the loop we can use
$$
\eqalign{
A^{[0]}   &= A^{N=0} \cr
A^{[1/2]} &= A^{N=1}-A^{[0]} \cr
A^{[1]}   &= A^{N=4}-4A^{N=1}+A^{[0]} \cr
A^{[3/2]} &= A^{N=6}-6A^{N=4}+9A^{N=1}-A^{[0]} \cr
A^{[2]}   &= A^{N=8}-8A^{N=6}+20A^{N=4}-16A^{N=1}+A^{[0]} \cr
}
\eqn\InverseDecomp
$$

\section{Four-graviton amplitudes}

We now present the one-loop 4-graviton results for all choices of
helicity.  We will give each result in the supersymmetric decomposition
form as described in the previous section.  We also quote the pure
gravity results explicitly.

The first step is to insert spinor helicity simplifications into the
kinematic expression (\use\MasterKin).  The spinor helicity method for
gravitons [\use\Berends,\use\SpinorGravity] is related to that for vectors
[\use\XZC] by
$$
\pol^{++} = \pol^+ \bar{\pol}^+ ,
\hskip 2cm \pol^{--} = \pol^- \bar{\pol}^-
$$
where $\pol^{\pm\pm}$ are the graviton helicity polarizations and
$\pol^{\pm}$ are the vector helicity polarizations defined by Xu,
Zhang and Chang.
We use the notation for spinor inner products
$\langle k_1^- | k_2^+ \rangle = \langle 1 2 \rangle$ and
$\langle k_1^+ | k_2^- \rangle = \spb1.2 $. The use of spinor helicity
techniques has proved extremely useful in QCD calculation.
All states are taken to be outgoing and may have plus or minus helicity.
There is no concept of colour ordering which is found in QCD amplitudes.
There are thus three independent helicity configurations for the four
point amplitude,  $(+,+,+,+)$, $(-,+,+,+)$ and $(-,-,+,+)$, the others being
obtained by conjugation from these.

For the $(-,+,+,+)$ and
$(+,+,+,+)$ one-loop amplitudes all the
supersymmetric components in the decomposition
vanish due to supersymmetric Ward identities
analogous to the situation in QCD [\use\Susy].
The tree level graviton amplitudes vanish for these
helicity configurations and hence the one-loop results are the leading order
for these configurations and have a simple form rather analogous to
a tree amplitude without logarithms or infinities. From the inverse
decomposition (\InverseDecomp) this amplitude for any particle content
is just proportional to the scalar contribution. The scalar contributions
to graviton scattering have been previously calculated
in ref.~[\Zak] but not in a spinor helicity basis. The results from the
string-based rules agree with these results and we have checked explicitly
that the cycle contributions cancel, demonstrating the Ward identities.
We find
$$
A(1^-,2^+,3^+,4^+) = N_s  {i\kappa^4\over(4\pi)^2}
\Bigl({s t \over u}
\Bigr)^2\Bigl({ \spb2.4^2 \over \spb1.2 \spa2.3 \spa3.4\spb4.1} \Bigr)^2
{(  s^2 +st +t^2 ) \over 5760 }
\anoneqn
$$
$$
A(1^+,2^+,3^+,4^+) =
- N_s {i\kappa^4\over(4\pi)^2}
\Bigl({s t \over \spa1.2\spa2.3\spa3.4\spa4.1} \Bigr)^2 {( s^2 +st +t^2 )
\over 1920  }
\eqn\FourPlus
$$
where
$$
N_s = N_B-N_F
$$
is the number of bosonic states in the loop minus the number of
fermionic states and $s=(k_1+k_2)^2$, $t=(k_1+k_4)^2$ and
$u=(k_1+k_3)^2$.  So, for instance, since a graviton is made up of two
helicity states the amplitudes for pure gravity are found by putting
$N_s=2$ in the above expressions.

For the $A(1^-,2^-,3^+,4^+)$ amplitude, none of the cycle
terms vanish, so these must be included.  We express their
contributions using the supersymmetric decomposition given in the
previous section.  The (complex) scalar amplitude is
$$
\eqalign{ A^{[0]}(1^-,2^-,3^+,4^+) =
+&{ {F\left (t-u\right )\left (t^{4}+9\,ut^{3}+46\,u^{2}t^{2}+9\,u^{3}t
+u^{4}\right )\ln (-t/-u)}\over 30\,s^{7} }
\cr
+&{ {F\left (2\,t^{4}+23\,ut^{3}+222\,u^{2}t^{2}+23\,u^{3}t+2\,u^{4}\right )}
\over 180\,s^{6} }
-{{Fu^{3}t^{3} (  \ln (-t/-u)^{2} +\pi^2)  }\over s^{8}}
\cr}
\anoneqn
$$
where $F$ is
$$
{i\kappa^4(4\pi)^{\eps}r_{\Gamma}\over 16(4\pi)^2}
\biggl( {st \spa1.2^4 \over \spa1.2\spa2.3\spa3.4\spa4.1 } \biggr)^2
=
{istu\kappa^2(4\pi)^{\eps}r_{\Gamma}\over 4(4\pi)^2} {\Atree(1^-,2^-,3^+,4^+)}.
\anoneqn
$$
$A^{tree}(1^-,2^-,3^+,4^+)$ is defined in eq.~(5.3) and
$$
\rg= { \Gamma^2(1-\eps)\Gamma(1+\eps) \over \Gamma(1-2\eps) }
\; .
\anoneqn
$$

The amplitudes for the supersymmetric multiplets given in
table~\TABb\ are
$$
\eqalign{
 A^{N=1} &=-{F( {t^{2}+14\,tu+u^{2}})\over{24\,s^{4}}}
+{F t^2
u^2\Bigl( \ln^2 ({\it -t/-u}) +\pi^2 \Bigr)
\over{2\,s^6}}
\cr
&\null \hskip 2.0 truecm
-{F{\left (
t-u\right )\left (t^{2}+8\,tu+u^{2}\right )\ln (-t/-u)}\over{12\,s^{5}}}
 \cr
 A^{N=4} &=  {F \over 2s^4} \bigg( {(t-u)s\ln(-t/-u)}
- {tu\bigl(\ln^2(-t/-u)+\pi^2\bigr)}
+ { s^2} \bigg)
 \cr
 A^{N=6} &={-F \over 2} \biggl( {\ln^2(-t/-u)+\pi^2 \over s^2} \biggr)
  \cr
 A^{N=8}  &= {2F \over \epsilon } \bigg( {\ln(-u) \over st}
+ {\ln(-t) \over su}
+ {\ln(-s) \over tu} \bigg)
\cr
&\null\hskip 1.0 truecm
+ {2F} \bigg( {\ln(-t)\ln(-s) \over st} + {\ln(-u)\ln(-t) \over tu}
+ {\ln(-s)\ln(-u) \over us} \bigg)
\cr}\anoneqn
$$
We chose to express the amplitude in the (unphysical) regime
where all momentum variables $s$, $t$ and $u$ are negative.
One can obtain expressions in the physical region by the substitution
$$
\ln (-s) \rightarrow \ln(|s|) -i\pi \Theta(s)
\anoneqn
$$
etc. $\Theta(s)$ is the Heavyside function where $\Theta(x)=1, x>0$
and $\Theta(x)=0, x<0$.

The pure gravity amplitude can be found using the expression in
eq.~(\use\InverseDecomp)
which gives the result
$$
\eqalign{
A^{[2]}(1^-,2^-,3^+,4^+) = F \biggl( &
 {2 \over \epsilon } \bigg(  {\ln(-u) \over st} + {\ln(-t) \over su}
+  {\ln(-s) \over tu}  \biggr)
\cr +&
{{2\,\ln (-u)\ln (-s)}\over{s
u}}
+{ {2\,\ln (-t)\ln (-u)}\over{tu}}
+{{2\,\ln (-t)\ln (-s)}\over{ts}}\cr
+&{ {\left (t+2\,u\right )\left (2\,t+u\right )\left (2\,t^{4}+2\,t
^{3}u-t^{2}u^{2}+2\,tu^{3}+2\,u^{4}\right )
(\ln^2 ({-t/-u})+\pi^2 ) }\over{s^{8}}}\cr
+&{ {\left (t-u\right )\left (
341\,t^{4}+1609\,t^{3}u+2566\,t^{2}u^{2}+1609\,tu^{3}+341\,u^{4}
\right )\ln (-t/-u)}\over{30\,s^{7}}}\cr
+&{{1922\,t^{4}+9143\,t^{3}u+14622\,t^{2}u^{2}+9143\,tu^{3}+1922\,u^{4}}\over
{180\,s^{6}}} \biggr)
}\anoneqn
$$
These expressions have the correct symmetry expected in the amplitude
under, for example, interchange of legs $3$ and $4$.

Only the $N=8$ amplitude has $1/\eps$ singularities. These are purely
IR singularities the result being UV finite. This is as expected since
the $N=8$ multiplet is the only multiplet containing gravitons
and the loop amplitudes with other particles circulating are
expected to be IR finite.

\section{Unitarity constraints \& consistency checks}

Unitarity, in the form of the Cutkosky rules,
is a strong constraint on amplitudes.
When cancellations occur in the loop momentum integrals it becomes
particularly restricting and if enough cancellations occur unitarity may
be enough to uniquely determine the amplitude.
Again, the supersymmetric decomposition is useful in this context since
cancellations occur within supersymmetry multiplets.
In ref~[\SusyOne] it was proved that, for a gauge theory amplitude, if the
$n$-point loop integral has at most $n-2$ powers of loop momentum in the
numerator the amplitude is uniquely determined from the cuts.
Unfortunately, since gravity amplitudes generally have $2n$ powers of
loop momentum this result is not as useful in determining amplitudes
since for large $n$ the power of the loop momentum polynomial
grows to be larger than $n-2$ even in the case of maximal cancellation
$N=8$. In this case we may apply this result, in principle,
for $2n-8 \leq n-2$ that
is for $n \leq 6$ to completely determine the amplitude.
Looking at table~\use\TABc\ we see that for the four point
amplitude this result may be applied to the $N=8$
and $N=6$ multiplet contributions. (Technically the result in [\use\SusyOne]
applies when the cancellations occur in the loop momentum polynomial rather
than in the Feynman parameter polynomial, however by examining the zero
mode integral, for example, in [\use\Mapping] one can see that a loop
momentum representation exists for the string based rules and the result
may be use.) In this section we will demonstrate how to apply unitarity to
obtain the cuts in the one-loop amplitudes. For the four point $N=8$ and $N=6$
contributions the results agree completely with the string based results and
for the remaining amplitudes the amplitudes are consistent with the cuts.

To calculate the cuts in all channels using the Cutkosky rules.
Consider the regime where one of the momentum invariants is positive
and the remainder are negative.  This allow us to isolate the cuts in
a single channel.  We can then use the Cutkosky rules to obtain
the cuts. To apply the cuts one needs
explicit, preferable compact, expressions for tree amplitudes.
In general, we consider the cut in the channel
$(k_{m_1}+k_{m_1+1}+\cdots+k_{m_2-1}+k_{m_2})^2$ for the
loop amplitude $A_{n;1}(1,2,\ldots,n)$, depicted in
\fig\CutsFigure\ and given by
$$
\eqalign{{i \over 2}
  &\int \dlips(-\ell_1,\ell_2)
  \ A^{\rm tree}(-\ell_1,m_1,\ldots,m_2,\ell_2)
  \ A^{\rm tree}(-\ell_2,m_2+1,\ldots,m_1-1,\ell_1). \cr
}
\eqn\CutEquation
$$
Instead of evaluating the phase-space integrals instead evaluate
the off-shell integral
$$
\eqalign{{i \over 2}
  &\int {d^{D} \ell_1 \over (2\pi)^D}
  \ A^{\rm tree}(-\ell_1,m_1,\ldots,m_2,\ell_2) {1\over \ell_2^2}
  \ A^{\rm tree}(-\ell_2,m_2+1,\ldots,m_1-1,\ell_1) {1\over \ell_1^2}
   \biggr|_{\rm cut} . \cr
}\eqn\caseacut
$$
whose cut is (\use\CutEquation).
This replacement is only valid in this channel. In evaluating this
off-shell integral, we may substitute
$\ell_1^2 = \ell_2^2 = 0$ in the numerator; any terms with $\ell_1^2$
or $\ell_2^2$ in the numerator cancels a cut propagator leading to an
integral without a cut in this channel.
Evaluating these cuts
requires the tree amplitudes for all possible intermediate
states, preferably in a compact form.
For the four point those  tree amplitudes
which have been calculated previously in refs.~[\use\Zak,\use\Sannan]
and in a helicity basis by Berends, Giele and
Kuijf in ref~[\use\Berends] are sufficient.
For pure gravity, the tree amplitudes $\Atree(-+++)$ and
$\Atree(++++)$ vanish but $\Atree(--++)$ is non-zero and is given by
$$
\Atree(1^-,2^-,3^+,4^+)=
{\kappa^2 \over 4}
\biggl( { \spa1.2^4 \over \spa1.2\spa2.3\spa3.4\spa4.1
 } \biggr)^2
\times { s t \over u }
\eqn\TreeResult
$$
We first note that the $\Aloop(-,+,+,+)$ and
$\Aloop(+,+,+,+)$ one loop amplitudes have
no logarithms and hence no cuts. This is consistent with the fact that
there are no non-vanishing
pairs of tree amplitudes which could contribute to the
Cutkosky rules.

To calculate the cuts in $\Aloop(1^-,2^-,3^+,4^+)$, first consider the cut
in the
$s$-channel, $(k_1+k_2)^2$ as shown in
\fig\FourCut a.
$$
{i \over 2}\int \dlips
\Atree(1^-,2^-,\ell_2^+,\ell_1^+) \times  \Atree( \ell_1^-,\ell_2^-, 3^+,4^+ )
\eqn\FourPtCut
$$
This is non-zero for the case where the
intermediate (cut) states are gravitons,
however when the intermediate states are otherwise the tree amplitudes are
zero. This is because the graviton vertex does not flip helicity of the
fermions [\use\Berends,\use\Flip]
hence the amplitude with two gravitons and two fermions
of the same helicity vanishes, $A(g,g,\psi^+,\psi^+)=0$.
Similarly the vector amplitudes and scalar amplitudes (with the
concept of helicity being replaced by particle/antiparticle) vanish.
Hence states other than gravitons do not contribute to
eq.~(\use\FourPtCut). For the supersymmetric decomposition this implies the
$s$-channel cut will only be non-zero for the $N=8$ contribution.
Inserting the graviton tree amplitudes into (\FourPtCut) yields,
$$
{i\kappa^4\over32}\int \dlips \biggl( { \spa1.2^4 \over \spa1.2\spa{2}.{\ell_2}
\spa{\ell_2}.{\ell_1}\spa{\ell_1}.1 } \biggr)^2
\times { s (k_2\cdot \ell_2) \over (k_1\cdot \ell_2) }
 \biggl( { \spa{\ell_1}.{\ell_2}^4 \over \spa3.4\spa4.{\ell_1}
\spa{\ell_1}.{\ell_2}
\spa{\ell_2}.3 } \biggr)^2
\times { s(k_3\cdot \ell_2) \over (k_4\cdot \ell_2) }
\anoneqn
$$
which we can rearrange, using the fact that $\ell_1$ and $\ell_2$
are onshell,
$$
{i\kappa^4\over32}s^2 \biggl( { \spa1.2^4 \over \spa1.2\spa3.4 } \biggr)^2
\int \dlips
{(k_2\cdot \ell_2) \over (k_1\cdot \ell_2) }{ (k_3\cdot \ell_2)
\over (k_4\cdot \ell_2) }
{ \spa{\ell_1}.{\ell_2}^4 \over \spa2.{\ell_2}^2 \spa{\ell_1}.1^2
\spa4.{\ell_1}^2\spa{\ell_2}.3^2 }
\anoneqn
$$
We can rearrange this, using
$$
\eqalign{
{\spa{\ell_1}.{\ell_2}^2\over \spa{\ell_1}.1^2 \spa2.{\ell_2}^2  }
& ={ \spa{\ell_1}.{\ell_2}^2 \spb1.2^2 \over  \spa{\ell_1}.1
\spa2.{\ell_2} \spa{\ell_1}.1 \spb1.2  \spb1.2 \spa{2}.{\ell_2} }
\cr &
={ \spa{\ell_1}.{\ell_2}^2 \spb1.2^2 \over
\spa{\ell_1}.1 \spa2.{\ell_2} \spa{\ell_1}.{\ell_2}\spb{\ell_2}.2
\spb1.{\ell_1} \spa{\ell_1}.{\ell_2}  }
\cr &
={ \spb1.2^2 \over
\spa2.{\ell_2}\spb{\ell_2}.2   \spb1.{\ell_1} \spa{\ell_1}.1 }
\cr &
={  \spb1.2^2 \over  4( k_2 \cdot \ell_2 )(k_1\cdot \ell_1 ) }
\cr}
\eqn\ArrangeA
$$
where we use
$\spa{\ell_1}.1 \spb{1}.2=-\spa{\ell_1}.{\ell_2}\spb{\ell_2}.2$
etc.
Similarly
$$
{\spa{\ell_1}.{\ell_2}^2\over  \spa4.{\ell_1}^2\spa{\ell_2}.3^2 }
={\spb3.4^2 \over  4( k_3  \cdot \ell_2 )(k_4\cdot \ell_1 ) }
\eqn\ArrangeB
$$
Noting that
$$
{ \spb1.2\spb3.4} ={ -s t \over \spa2.3\spa4.1 }
\anoneqn
$$
we obtain the form of the cut
$$
\eqalign{{i\kappa^2\over8}
A^{tree}(1^-,2^-,3^+,4^+) s^3 t u
\int d{\rm LIPS}
{(k_2\cdot \ell_2) \over (k_1\cdot \ell_2) }{ (k_3\cdot \ell_2)
\over (k_4\cdot \ell_2) }
\times
{ 1 \over 16( k_2 \cdot \ell_2 )(k_1\cdot \ell_1 )
( k_3  \cdot \ell_2 )(k_4\cdot \ell_1 ) }
\cr
=
{i\kappa^2\over8}
A^{tree}(1^-,2^-,3^+,4^+) s^3 t u
\int d{\rm LIPS}
{ 1 \over 16(k_1\cdot \ell_2)(k_4\cdot \ell_2)
            (k_1\cdot \ell_1 )(k_4\cdot \ell_1 ) }
\cr}
\eqn\CUTaa
$$
There is a useful identity, requiring the fact that both trees are
on-shell.
$$
\eqalign{
& {1 \over(k_1\cdot \ell_2)(k_4\cdot \ell_2)
(k_1\cdot \ell_1 )(k_4\cdot \ell_1 ) }
\cr &
={4 \over s^2 }
\biggl( {-1 \over (k_1\cdot \ell_1)(k_4\cdot \ell_1 )}
       +{1 \over (k_1\cdot \ell_2)(k_4\cdot \ell_1 )}
       +{1 \over (k_1\cdot \ell_1)(k_4\cdot \ell_2 )}
       +{-1 \over (k_1\cdot \ell_2)(k_4\cdot \ell_2 )}
\biggr)
\cr &
={16 \over s^2 }
\biggl( {1 \over (k_1-\ell_1)^2(k_4+\ell_1 )^2}
       +{1 \over (k_1+\ell_2)^2(k_4+\ell_1 )^2 }
       +{1 \over (k_1-\ell_1)^2(k_4-\ell_2 )^2 }
       +{1 \over (k_1+\ell_2)^2(k_4-\ell_2 )^2 }
\biggr)
\cr }
\anoneqn
$$
This uses momentum conservation extensively.  Inserting the two
propagators $1/\ell_1^2$ and $1/\ell_2^2$ and replacing
$\int\dlips$
by $\int d^D \ell/(2\pi)^D$ as in eq.~(\use\caseacut) the cut
in eq.~(\use\CUTaa) can now be
recognised as the cut of the sum of two scalar box integrals with
orderings $1234$ and $2134$ ( the four terms above only correspond to
two independent boxes.) These boxes have coefficients,
$$
2{\kappa^2\over8} \Atree(1^-,2^-,3^+,4^+) st u
\anoneqn
$$

Next, consider cuts in the channel $(k_1+k_4)^2$. These in
general are more complex and also depend upon the multiplet under
consideration.  As can be seen from fig.~\use\FourCut b in this case
all particles contribute.  To evaluate the cut one needs the four
point amplitudes with two external gravitons and two scalars or
fermions or vectors or gravitinos. These may be obtained from the four
graviton amplitudes using an extended form of the supersymmetric ward
identities [\use\Susy]. From these, we obtain
$$
\eqalign{
A(g^-,\phi^-,\phi^+,g^+) & = { \spa1.3^4 \over \spa1.2^4 }
A(g^-,g^-,g^+,g^+)
\cr
A(g^-,\Lambda^-,\Lambda^+,g^+) & = { \spa1.3^3\over \spa1.2^3 }
A(g^-,g^-,g^+,g^+)
\cr
A(g^-,A^-,A^+,g^+) & = { \spa1.3^2 \over \spa1.2^2 }
A(g^-,g^-,g^+,g^+)
\cr
A(g^-,\psi^-,\psi^+,g^+) & = { \spa1.3\over \spa1.2 }
A(g^-,g^-,g^+,g^+)
\cr}
\eqn\Relative
$$
where $(g,\psi,A,\Lambda,\phi)$ are the members of the
$N=8$ multiplet.

It is useful to count each states contribution to the cut relative to that for
a scalar.
All these states will contribute to the cut,
$$
{i \over 2} \int \dlips A(4^+,1^-,\phi(\ell_2)^-,\phi(\ell_1)^+)
\times  A( \phi(\ell_1)^-,\phi(\ell_2)^+, 2^-,3^+ )
\times \rho_{N=8}
\anoneqn
$$
which is explicitly
$$
{i\kappa^4\over32}{ t^2 \over \spa2.3^2\spa4.1^2 }
\int\dlips
{ \spa1.{\ell_1}^4\spa2.{\ell_1}^4\spa1.{\ell_2}^2\spa2.{\ell_2}^2
\over \spa3.{\ell_1}^2  \spa4.{\ell_1}^2 \spa{\ell_1}.{\ell_2}^4  }
{ (k_1\cdot \ell_2)(k_2\cdot \ell_2) \over
(k_1\cdot \ell_1)(k_2\cdot \ell_1) }
\rho_{N=8}
\anoneqn
$$
The factor $\rho$ for $N=8$ will be
$$
\rho_{N=8}=
x^8 -8x^6 +28x^4 -56x^2 +70 -56x^{-2}+28x^{-4}-8*x^{-6} +x^{-8}
\anoneqn
$$
where
$$
x^2 = { \spa1.{\ell_2}\spa2.{\ell_1} \over \spa1.{\ell_1}\spa2.{\ell_2} }
\anoneqn
$$
The central term, $70$, arises from the $35$ complex scalars, the $x^2$
and $x^{-2}$ from the Weyl fermions, the $x^4$ and $x^{-4}$ from the
vectors, the $x^6$ and $x^{-6}$ from the gravitinos and the $x^8$
and $x^{-8}$ from the gravitons. These relative weights compared to the scalar
contribution are obtained from eq.~(\use\Relative).
This simplifies to
$$
\eqalign{
\rho_{N=8} = (x-x^{-1})^8 = { (x^2-1)^8 \over x^8 }
& ={ (\spa1.{\ell_2} \spa2.{\ell_1}-\spa1.{\ell_1}\spa2.{\ell_2} )^8
\over \spa1.{\ell_2}\spa2.{\ell_1}\spa1.{\ell_1}\spa2.{\ell_2})^4 }
\cr
& ={\spa1.2^8 \spa{\ell_1}.{\ell_2}^8
\over (\spa1.{\ell_2}\spa2.{\ell_1}\spa1.{\ell_1}\spa2.{\ell_2})^4 }
\cr}
\anoneqn
$$
using the identity $\spa{a}.b\spa{c}.d=\spa{a}.c\spa{b}.d+\spa{a}.d\spa{c}.b$.
The cut then becomes
$$
 {i\kappa^4\over32}t^2{ \spa1.2^8 \over \spa2.3^2\spa4.1^2 }
\int\dlips
{ (k_1\cdot \ell_2) \over (k_1\cdot \ell_1) }
{ (k_2\cdot \ell_2) \over (k_2\cdot \ell_1) }
{
\spa{\ell_1}.{\ell_2}^4 \over
             ( \spa1.{\ell_2}\spa{\ell_1}.4\spa{\ell_2}.2\spa3.{\ell_1} )^2 }
\anoneqn
$$
Similarly to before we have the identities
$$
\eqalign{
{\spa{\ell_1}.{\ell_2}^2\over \spa{\ell_1}.4^2 \spa1.{\ell_2}^2  }
&={  \spb4.1^2 \over  4( k_1 \cdot \ell_2 )(k_4\cdot \ell_1 ) }
;
\hskip 1.0 cm
{\spa{\ell_1}.{\ell_2}^2\over  \spa3.{\ell_1}^2\spa{\ell_2}.2^2 }
={\spb2.3^2 \over  4( k_2  \cdot \ell_2 )(k_3\cdot \ell_1 ) }
\cr}
\anoneqn
$$
and
$$
{ \spb4.1^2\spb2.3^2 } =
{ t^2 s^2 \over \spa1.2^2 \spa3.4^2 }
\anoneqn
$$
which give the cut to be
$$
 {i\kappa^4\over32}t^4 s^2  \biggl(
{ \spa1.2^8 \over\spa1.2^2\spa2.3^2\spa3.4^2 \spa4.1^2 } \biggr)
\int\dlips
{ (k_1\cdot \ell_2) \over (k_1\cdot \ell_1) }
{ (k_2\cdot \ell_2) \over (k_2\cdot \ell_1) }
\times
{ 1 \over 16 ( k_1 \cdot \ell_2 )(k_4\cdot \ell_1 )
             ( k_2  \cdot \ell_2 )(k_3\cdot \ell_1 ) }
\anoneqn
$$
which is
$$
{i\kappa^2\over8}A^{tree}(4^+,1^-,2^-,3^+) t^2 s u
\int\dlips { 1 \over 16 (k_1\cdot \ell_1)(k_2\cdot \ell_1)
                  (k_4\cdot \ell_1 )(k_3\cdot \ell_1 ) }
\anoneqn
$$
This is just the analogue of the previous case. We
can carry out the same factorization as before and find the two
appropriate boxes with ordering 1234 and 1243.
The cuts in the $(k_1+k_3)^2$ channel are obtained exactly as the
$t$-channel cuts.
Replacing, $\int \dlips$ by
$\int d^{D}\ell / (2\pi)^2$ we can thus deduce
that the cuts in all channels
are described by the sum over the scalar boxes
with coefficients
$i\kappa^2 \Atree stu/4$.
The scalar box, for ordering of legs $1234$ is
$$
\eqalign{
I_4& =
{ \rg (4\pi)^{\eps} \over (4\pi)^2 }
{ 1 \over s t} \biggl\{
{2\over \eps} \Bigl[ (-s)^{-\eps} +(-t)^{-\eps} \Bigr]
-\ln^2(-s/-t) -\pi^2 \biggr\}
\cr & =
{ \rg (4\pi)^{\eps} \over (4\pi)^2 }
{ 1 \over s t} \biggl\{ {4 \over \eps^2}
-{2\over \eps}\Bigl( \ln(-s)+\ln(-t) \Bigr)
+2\ln(-s)\ln(-t) -\pi^2 \biggr\}
\cr}
\anoneqn
$$
This sum over boxes then evaluates to
$$
\eqalign{
A^{N=8}(&1^-,2^-,3^+,4^+) =
{i\kappa^2 }
{ \rg (4\pi)^{\eps} \over (4\pi)^2 }
\Atree(1^-,2^-,3^+,4^+)
\cr
\times &\biggl(
{ 2s \ln(-s)+2t\ln(-t)+2u\ln(-u) \over \eps }
+2s\ln(-t)\ln(-u)+2t\ln(-u)\ln(-s)
+2u\ln(-s)\ln(-t)
\biggr)
\cr}
\anoneqn
$$
Since we have produced an expression with the correct cuts which is
written in terms of integral functions we can use the results of
ref.~[\SusyFour,\SusyOne] to deduce that this expression is the
entire amplitude.
This is in agreement with the explicit calculation of the previous section.

For the other cases, the, $N=6$,$N=4$,$N=2$ and $N=0$ contributions,
we may perform a similar calculation to the $N=8$ case
but with differing $\rho$.
We will have
$$
\eqalign{
\rho_{N=6}= (x-x^{-1} )^6 =
{ \spa1.2^6 \spa{\ell_1}.{\ell_2}^6 \over
(\spa1.{\ell_2}\spa2.{\ell_1}\spa1.{\ell_1}\spa2.{\ell_2})^3 }
\cr
\rho_{N=4}= (x-x^{-1} )^4 =
{ \spa1.2^4 \spa{\ell_1}.{\ell_2}^4 \over
(\spa1.{\ell_2}\spa2.{\ell_1}\spa1.{\ell_1}\spa2.{\ell_2})^2 }
\cr
\rho_{N=2}= (x-x^{-1} )^2 =
{ \spa1.2^2 \spa{\ell_1}.{\ell_2}^2 \over
(\spa1.{\ell_2}\spa2.{\ell_1}\spa1.{\ell_1}\spa2.{\ell_2}) }
\cr}
\anoneqn
$$
A similar calculation of the cuts may be performed. For the $N=6$
calculation the cuts are enough to completely reconstruct the amplitude and we
obtain an amplitude in agreement with the explicit string based-calculation.
For the other case, the cuts would not be enough to completely specify the
amplitude but provide strong consistency checks on the amplitudes.
In general, the cuts are relatively simple to calculate when compact forms
for the tree amplitudes exist and where the cuts specify the amplitude
completely they would be the calculational method of choice.

In the next section we will illustrate the use of the cuts to obtain
the logarithmic parts of two-loop amplitude $A(+,+,+,+)$.


\section{The Cuts in $A^{\rm 2-loop}(+,+,+,+)$ }

In this section we will use the Cutkosky rules to calculate the cuts
in the two-loop pure gravity amplitude $A^{\rm 2-loop}(+,+,+,+)$ . It is
possible to do this because of the simple form of the one-loop
amplitude $A^{\rm 1-loop}(+,+,+,+)$. If we consider the
two-loop cuts we must consider the cuts in the three-particle
intermediate states as shown in \fig\FIGTwoLoop a. Fortunately, these
cuts vanish because one of the two tree amplitudes must inevitably
have a single negative helicity and this tree vanishes. Considering the
remaining possibilities we find, for example in the $s$-channel that
the configurations in
figs.~\FIGTwoLoop b\ and
\FIGTwoLoop c\  may contribute. In this case we have a product of
$A^{\rm tree}(-,-,+,+)$ and
$A^{\rm 1-loop}(+,+,+,+)$ neither of which vanish.
Since the one-loop amplitude does not
contain logarithms or dilogarithms the evaluation of the cut is analogous to
calculating a one-loop cut and we are able to do so.
Explicitly the $s$-channel cut in fig.~\FIGTwoLoop b\ is
$$
\hskip -.4 cm
\eqalign{{i \over 2}
  &\int \dlips(-\ell_1,\ell_2)
  \ A^{\rm 1-loop}(1^+,2^+,-\ell_1^+,\ell_2^+)
  \ A^{\rm tree}(-\ell_2^-,\ell_1^-,3^+,4^+), \cr}
\anoneqn
$$
where $\dlips(-\ell_1,\ell_2)$ denotes the Lorentz-invariant phase
space measure. The explicit form of the tree and one-loop amplitudes
is obtainable from eq.~(\use\FourPlus) and eq.~(\use\TreeResult).
With the parameterisation shown
$$
\eqalign{
A^{\rm 1-loop}(1^+,2^+,-\ell_1^+,\ell_2^+)
&={-2i\kappa^4 \over (4\pi)^2 }
{1\over1920 }\biggl( { s (2k_2 \cdot \ell_2)
\over \spa1.2\spa2.{\ell_2}\spa{\ell_2}.{\ell_1}
\spa{\ell_1}.1 } \biggr)^2
\times \Bigl( s^2 +s(2k_2 \cdot \ell_2) +(2k_2 \cdot \ell_2)^2 \Bigr)
\cr
A^{\rm tree}(-\ell_2^-,\ell_1^-,3^+,4^+)
&={i\kappa^2 \over 4} \biggl({ \spa{\ell_2}.{\ell_1}^3
\over \spa{\ell_2}.3 \spa3.4 \spa4.{\ell_1} }
\biggr)^2 { s (2k_3 \cdot \ell_2) \over (2k_4\cdot \ell_2) }
\cr}
\anoneqn
$$

The evaluation of this cut employs many of the tricks already
used to obtain checks of the cuts for the one-loop results.
Using eq.~(\use\ArrangeA)\ and (\use\ArrangeB) we can simplify the cut to
$$
\sim
{1\over1920}{ s^5 t^2 \over( \spa1.2 \spa2.3\spa3.4 \spa4.1 )^2 }
\int\dlips
{ (2k_2 \cdot \ell_2 )
\Bigl( (s^2+s(2k_2 \cdot \ell_2)+(2k_2 \cdot \ell_2)^2 \Bigl)
\over
(2k_1\cdot \ell_1 )(2k_4\cdot \ell_1) (2k_4\cdot \ell_2 ) }
\anoneqn
$$
Using
$$
{1 \over (2k_4\cdot \ell_1) (2k_4\cdot \ell_2 ) }
={1 \over s } ( {1 \over (2k_4\cdot \ell_1) }-{1 \over(2 k_4\cdot \ell_2) } )
={1 \over s } ( {1 \over (k_4+ \ell_1)^2 }+{1 \over (k_3+ \ell_1)^2 } )
\anoneqn
$$
this reduces to a sum of two box integrals. However since
$(2k_2 \cdot \ell_2 )=-(2k_1\cdot \ell_1)$
the box integrands further reduce to the triangle integrals shown in
\fig\FIGtwoloopRed\ with loop momentum polynomial
$$
\Bigl( s^2+s(2k_2 \cdot \ell_2)+(2k_2 \cdot \ell_2)^2 \Bigl)
\eqn\PolynomTri
$$
Evaluating this integral gives the result,
$$
\Bigl(
{ s^2 +t^2 +u^2 \over 2 \eps^2 }
+{1 \over 2 \eps }( 3u^2+3t^2-2s^2 )
+{1\over 4}(7u^2+7t^2-5s^2)
\Bigr)\times (-s)^{-1-\eps}
\anoneqn
$$
The second triangle gives the same result and the contribution from
fig.~\FIGTwoLoop c\ is also the same. This equation has the correct
cut in the $s$-channel. The $t$ and $u$ channel cuts are obtained from
this by permuting $s$,$t$ and $u$.  We can thus obtain an expression
for the logarithmic parts of $A^{\rm 2-loop}(1^+,2^+,3^+,4^+)$
$$
\eqalign{
A^{\rm 2-loop}(1^+,2^+,& 3^+,4^+)
\sim
A^{\rm 1-loop}(1^+,2^+,3^+,4^+) \times
\cr
\biggl(
&
{ (-s)^{1-\eps}+(-t)^{1-\eps}+(-u)^{1-\eps} \over 2 \eps^2 }
+{1 \over 2  }
{ ( 3u^2+3t^2-2s^2 ) \over ( s^2 +t^2 +u^2 ) }
{ (-s)^{1-\eps} \over  \eps }
\cr
+&{1 \over 2  }
{ ( 3u^2+3s^2-2t^2 )  \over ( s^2 +t^2 +u^2 ) }
{ (-t)^{1-\eps} \over  \eps }
+{1 \over 2  }
{ ( 3s^2+3t^2-2u^2 )  \over ( s^2 +t^2 +u^2 ) }
{ (-u)^{1-\eps} \over  \eps }
\cr
&
+{\rm polynomials }
\biggr)
\cr}
\anoneqn
$$
where the polynomial pieces are not obtainable from the cuts.
Associated with the logarithms in the above expression are $1/\eps$
poles. The $A(++++)$ two-loop amplitude is an interesting object in
perturbative gravity. Gravity is non-renormalisable at two loops
however not all amplitudes contain obvious non-renormalisable
UV infinities. As is well known [\use\NieWu],
the two loop amplitude $A(--++)$
does not contain such infinities and the infinities reside
in the so-called ``helicity-flip'' amplitudes $A(++++)$ and
$A(-+++)$. Calculation of the UV infinities for these amplitudes
is a considerable undertaking and the explicit verification of
the non-renormalisability of gravity at two-loops was an important
but difficult calculation [\use\GoroffSagnotti,\use\VanDVen].
Although we have a suggested form for the infinities in this amplitude,
obtained with an almost trivial calculation, we are not able to extract
the UV infinity. Specifically the recognising of the UV and IR infinities
is not possible by examining the cut integrals. When the cut was reduced
to a triangle integral with polynomial in eq.~(\use\PolynomTri)
one would normally extract the UV infinity from the $\ell_{\mu}\ell_{\nu}$
which would give
$$
  { 1\over \eps } \times \delta_{\mu\nu}
\anoneqn
$$
However this yields zero since the coefficient is
$k_1^{\mu}k_1^{\nu}$.  Since the cut is only sensitive up to terms
proportional to $\ell_1^2$ and $\ell^2_2$ the loop momentum polynomial
could have been replaced by
$$
\Bigl( (2k_1 \cdot \ell_2)^2+(2k_1 \cdot \ell_2)(2k_2 \cdot \ell_2)
+(2k_2 \cdot \ell_2)^2 \Bigl)
\eqn\PolynomTriB
$$
without affecting the cut. However this polynomial would give an
UV infinity since the $\delta$-function would no longer vanish
for the middle term. Thus although tempting, we are unable to deduce
the coefficients of the UV infinities in the two-loop
amplitude, at least without further information.

For the configuration $A(1^-,2^+,3^+,4^+)$ the form of the one-loop
amplitude also does not contain logarithms and one may evaluate the equivalent
cut diagrams to those in fig.~\FIGTwoLoop b and \FIGTwoLoop c
however in this case there is a diagram as depicted in
\fig\FIGBadGuy\
which is non-zero and needs genuine two-loop integrals to
evaluate.

Two-loop amplitudes are formidable calculations in general.
Progress towards a string based or string inspired method has been
made both based upon the infinite string tension limit [\Kaj] and
upon the world-line formalism [\Wline]. In any formalism
we expect unitarity to provide very useful checks upon the calculations
as evidenced by the simplicity by which we obtained the cuts in
$A^{\rm 2-loop}(++++)$
(albeit the simplest  case possible).


\section{Conclusions}
Recently new rules for calculations of one-loop amplitudes have been
constructed using string theory methods.  These have been
successfully used to do calculations in both gauge theory and gravity
which have not been practical using conventional methods.  In this
paper we gave a detailed description of a set of rules for gravity
derived from string theory.  We used the rules to calculate one-loop
amplitudes for four graviton scattering.  These covered theories
with arbitrary particles content.  Our results are consistent both with
previous calculations and formal arguments concerning divergences.
Although perturbative quantum gravity is a non-renormalisable field theory
the UV divergences do not appear in our calculations.

In order to simplify the calculations we used a supersymmetric decomposition
of the amplitudes also inspired by string theory.  In a supersymmetric
theory,
provided one uses a suitable formalism, there
are generally a large number of cancellations between different particle
amplitudes.  Careful choice of the supersymmetric multiplet amplitudes
calculated enabled us to exploit these simplifications.  Individual
particle contributions could then be found from linear combinations of these
supersymmetric amplitudes. The supersymmetric
decomposition in the string based rules has proved extremely useful
since it reduces the degree of the momentum loop polynomial diagram by
diagram. Such cancellations do not occur even in a normal superfield
formalism but are familiar if one uses a background field superfield
formalism. The calculational advantages of calculating
$S$-matrix elements using background field methods are widespread
especially in situations where many pure gauge vertices appear or in
situations where cancellations are possible.

We used unitarity constraints to check the amplitudes calculated.
These constraints were found by use of the Cutkosky rules.  For the
$N=8$ and $N=6$ multiplets unitarity determined the amplitudes
completely.  For the remaining cases those parts of the amplitudes
containing cuts could be checked.  Since the loop momentum polynomial
grows as $2n$ for an $n$-point amplitude rather than as $n$ for gauge
theories the Cutkosky rules are not as powerful a constraint in
perturbative gravity as in gauge theories.

We also showed how the Cutkosky method could be used to obtain the
logarithmic parts of a two loop gravity amplitude by calculating the
cuts for the two-loop $(+,+,+,+)$ helicity amplitude. This amusing
calculation is incomplete since there are potential polynomial terms
but it does illustrate the potency of unitarity.

In conclusion we have found considerable calculational benefits from
calculating using the string-based methods [\use\StringBased,\PASCOS] and
derivatives thereof [\use\SusyFour,\use\SusyOne].
Many of the techniques motivated
by string theory have proved useful in these gravity calculations and
we expect these to have wider validity.

{\it Acknowledgments} We thank Zvi Bern, Lance Dixon,
Joe Minahan and Tokuzo Shimada for useful
discussions. This work was supported by the NSF under grant
PHY-9218990 (DCD), by the Department of Energy under grant
DE-FG03-91ER40662 (DCD) and by PPARC
(PSN). One of us (PSN) would like to thank
UCLA for hospitality.

\listrefs

\vfill\eject

\centerline{\bf Figure Captions}

\use\CutsFigure\ {\it A generic cut in the amplitude is given by an integral
over lorentz-invariant phase space (LIPS) of the product of two tree
amplitudes.}

\vskip 0.4 cm

\use\FourCut\  {\it The helicity configurations for the $s$ and $t$ cuts in
$A^{1-loop}(+,+,+,+)$ }

\vskip 0.4 cm

\use\FIGTwoLoop\  {\it The three possible contributions to
the $s$-channel cut for $A^{2-loop}(+,+,+,+)$. Contribution (a)
is vanishing since there is no choice of helicity for the
intermediate legs where both trees are non-vanishing.}

\vskip 0.4 cm

\FIGtwoloopRed\  {\it After manipulations, the $s$-channel cut in
$A^{2-loop}(+,+,+,+)$ reduces to a sum of simple triangle integrals
with integrands quadratic in the loop momentum.}

\vskip 0.4 cm

\FIGBadGuy\ {\it This helicity configuration is non-vanishing and
contributes to the cuts in $A^{2-loop}(-,+,+,+)$. }

\bye